\documentclass[aps,prd,onecolumn,nofootinbib,superscriptaddress]{revtex4}
\pdfoutput=1
\usepackage{graphicx}%Include figure files
\usepackage{amsmath}
\usepackage{amsfonts}
\usepackage{amssymb}
\usepackage{color}%
\usepackage{dcolumn}% Align table columns on decimal pointl
\usepackage{braket}
\usepackage{eurosym}
\usepackage[usenames,dvipsnames,svgnames]{xcolor}

\newcommand{\RNum}[1]{\uppercase\expandafter{\romannumeral #1\relax}}

\selectfont

\usepackage[title]{appendix}
\usepackage{wrapfig,boxedminipage,subfigure,epsfig}

\usepackage{ulem}   % a package related to strikethrough
\usepackage[english]{babel}

\usepackage{multirow}
\usepackage{array}
\usepackage{booktabs}
\usepackage{adjustbox}
\usepackage{caption}
\usepackage[colorlinks=true,linkcolor=blue,urlcolor=blue,filecolor=black,citecolor=red,pdfstartview=FitV,pdftitle={},pdfsubject={},pdfkeywords={},pdfpagemode=None,bookmarksopen=true]{hyperref}

\begin{document}
\thispagestyle{empty}
% \preprint{\hfill {\small {ICTS-USTC/PCFT-24-03}}}
%<<<<<<<<<<<<< TITLE >>>>>>>>>>>>>>>%
\title{Quasinormal mode families: classification and competition}
	
%<<<<<<<<<<<<< AUTHOR >>>>>>>>>>>>>>>%

\author{Zhen-Hao Yang}
\email{yangzhenhao\_yzu@163.com}
\affiliation{Center for Gravitation and Cosmology, College of Physical Science and Technology, Yangzhou University, Yangzhou 225002, China}

\author{Liang-Bi Wu}
\email{liangbi@mail.ustc.edu.cn}
\affiliation{School of Fundamental Physics and Mathematical Sciences, Hangzhou Institute for Advanced Study, UCAS, Hangzhou 310024, China}

\author{Xiao-Mei Kuang}
\email{xmeikuang@yzu.edu.cn}
\affiliation{Center for Gravitation and Cosmology, College of Physical Science and Technology, Yangzhou University, Yangzhou 225002, China}
	
\author{Wei-Liang Qian}	
\email{wlqian@usp.br}
\affiliation{Escola de Engenharia de Lorena, Universidade de S\~ao Paulo, 12602-810, Lorena, SP, Brazil}
\affiliation{Center for Gravitation and Cosmology, College of Physical Science and Technology, Yangzhou University, Yangzhou 225002, China}	
%<<<<<<<<<<<<< DATE >>>>>>>>>>>>>>>%
\date{\today}
	
%======================================%
%<<<<<<<<<<<<< ABSTRACT >>>>>>>>>>>>>>>%
%======================================%
\begin{abstract}
The perturbation spectra of black holes {beyond standard vacuum
black hole solutions within} general relativity (GR) may exhibit complex structures with long-lived modes. This usually generates echo-like modulations on the ringdown signal, {which typically originate from modified boundary conditions associated with exotic compact objects. Recent studies also reveal that they can instead be arised from the multi-peaked structure of the perturbation potential.} {However, while some case-by-case studies have been carried out,} a framework for understanding the internal structure of such spectra, the physical nature of different mode families, and their dynamical excitation {remains to be fully systematized.} In this paper, we address this issue by proposing a potential methodology that combines frequency-domain classification with time-domain analysis, using a hairy Schwarzschild black hole that admits a double-peak perturbative potential as a theoretical platform
Our analysis of the quasinormal mode (QNM) spectrum identifies two distinct families of modes: the photon sphere (PS) family, arising from delocalized scattering resonances, and the echo family, corresponding to highly localized quasi-bound states. We then develop a windowed energy analysis framework in the time domain,  which discloses a dynamic competition for dominance between these families. In particular, our results explicitly show that this competition is sensitive to the properties of the initial perturbation source, and that higher-overtone echo modes can dominate in the observed signal, which are in contrast to the standard PS mode in GR. This study establishes the dynamic evolution of this energy competition as a new observational signature for probing new physics and further motivate a supplemental framework for analyzing long-lived ringdown signals.
\end{abstract}

\maketitle
\tableofcontents
	
%======================================%
%<<<<<<<<<<< Introduction >>>>>>>>>>>>>%
%======================================%	
\section{Introduction}
The dawn of the gravitational-wave (GW) astronomy era, inaugurated by the landmark observations of the {Laser Interferometer Gravitational-Wave Observatory (LIGO)} collaborations~\cite{LIGOScientific:2016aoc}, has opened an unprecedented window into the strong field regime of gravity. Future space-based detectors such as {Laser Interferometer Space Antenna (LISA)} \cite{Finn:2000sy}, Taiji \cite{Hu:2017mde}, and TianQin \cite{TianQin:2015yph} and more promise to expand this window even further. One important scenario in this field is black hole spectroscopy~\cite{Berti:2025hly,Berti:2005ys}, the idea that by analyzing the ``ringdown'' signal of a post-merger black hole, one can probe its intrinsic properties analogous to how atomic spectroscopy reveals the structure of atoms. Within the standard paradigm, the ringdown waveform is modeled as a superposition of QNMs  \cite{Vishveshwara:1970zz, Kokkotas:1999bd, Berti:2009kk, Konoplya:2011qq}. These modes, characterized by their complex frequencies, are considered carriers of information about the final black hole's mass and spin, providing a powerful framework to test the foundational no-hair theorem of GR \cite{Berti:2005ys, Dreyer:2003bv, Gossan:2011ha}.  {Recent applications of black hole spectroscopy have been extended beyond the standard paradigm to incorporate nonlinearities \cite{Gleiser:1995gx, Cheung:2023vki, Redondo-Yuste:2023ipg, Zhu:2024dyl}, overtone contributions \cite{Giesler:2019uxc, Bhagwat:2019dtm, Baibhav:2023clw}, and environmental effects \cite{Barausse:2014tra, Cheung:2021bol, Biswas:2023ofz}. Furthermore, the intrinsic structure of the spectrum has been shown to exhibit rich features, including spectral instabilities \cite{Nollert:1996rf, Jaramillo:2020tuu, Cao:2024oud, Cai:2025irl} and resonant behaviors such as avoided crossings and exceptional points \cite{Onozawa:1996ux,Cook:2014cta, Motohashi:2024fwt,Lo:2025njp,Yang:2025dbn,Cao:2025afs,PanossoMacedo:2025xnf}. Moreover, the scope of ringdown analysis has broadened to probe fundamental physics, such as spacetime symmetries \cite{Ghosh:2024het}, non-gravitational modes arising from extra degrees of freedom \cite{Crescimbeni:2024sam}, and so on.}

Within this expanding landscape, recent theoretical explorations have shown that the emergence of new QNM families is a widespread phenomenon in a variety of physical scenarios. These new  families can be categorized by their physical origins. Firstly, modifications to spacetime's asymptotic or topological structure can introduce new QNM families. A prime example is found in \cite{Cardoso:2017soq}, where, in addition to the familiar photon sphere (PS) family tied to the black hole's light ring, a distinct and long-lived de Sitter family arises from the presence of the cosmological horizon. Exotic compact objects like wormholes \cite{Cardoso:2016rao,Bueno:2017hyj,Qian:2025lau}, or even specific braneworld models \cite{Tan:2024qij}, can feature effective potential wells that trap perturbations, giving rise to long-lived echo families \cite{Cardoso:2019rvt}. Additionally, the echo modes were also found and analyzed by imposing {a modified boundary condition at the horizon \cite{Abedi:2020ujo, Cardoso:2017cqb, Cardoso:2016oxy} or} a non-smooth effective potential \cite{Liu:2021aqh,Daghigh:2025wcw,Xie:2025jbr}. Secondly, the coupling between gravitational perturbations and other matter fields invariably leads to mixed spectra. For examples,  the gravito-electromagnetic coupling in Kerr-Newman spacetimes splits the QNM spectra into PS and near-horizon families \cite{Dias:2015wqa}; and the gravitational and scalar modes in various scalar-tensor theories become intertwined \cite{Guo:2022rms, Blazquez-Salcedo:2016enn,Lin:2025lsz}.
The coupling within Einstein-Maxwell-scalar theories can also generate long-lived modes and associated echoes, a notable feature that has been shown to persist even at the fully nonlinear level \cite{Sang:2022hng,Melis:2024kfr}.
Lastly, the intrinsic properties of the perturbing fields themselves, such as the mass of a Proca field, can generate quasi-bound states that form their own spectral families \cite{Rosa:2011my,Lei:2023wlt}. All these studies suggest that complex spacetime geometries or matter contents could leave their imprints directly on the global structure of the QNM spectra.

Since the perturbation spectrum can exhibit such rich structures, it is significant to take care of the global structure of the spectra besides the QNM spectra themselves. Especially, the spectra structure itself, including the number of mode families, their physical origins, hierarchical arrangement, and interaction, constitutes a new and largely untapped channel of information about the spacetime geometry and the underlying physical theory. Cardoso \textit{et al}. have laid the groundwork by performing a physical classification of QNM families in the frequency domain \cite{Cardoso:2017soq}, however, the dynamical picture of how these distinct families are excited and compete for dominance as the time goes by remains an open issue. This is important because the temporal evolution of the spectral structure is very crucial for a complete extraction of information from the ringdown signal. Our study aims to bridge this gap by proposing a preliminary but systematic methodology for analyzing complex systems that possesses multiple QNM families. Encouraged by the above, we select an theoretical testbed, a hairy Schwarzschild black hole constructed via the gravitational decoupling (GD) method \cite{Ovalle:2020kpd}, which naturally gives rise to a double-peak effective potential capable of supporting both PS and echo spectrum families.

Our methodology consists of a two-step workflow. The first step is a physical classification of modes in the frequency domain, which is made by WKB approximation, and we verify the reliability of the results of the QNM spectra with a set of methods. The second step is a dynamical analysis in time domain. To this end, we compute the complex amplitude of the master scalar \cite{Berti:2007dg, Nollert:1998ys} and develop a \textit{windowed energy analysis framework}. By calculating the energy fraction of each mode as a function of parameters related with a sliding observation window, we construct a dynamic portrait of how different families compete for dominance and reveal the sensitivity of this competition to the initial perturbation source. {It is noted that our proposed scheme can be applied for exploring QNM families in a broader studies associated with multi-peaked potentials. Here
our analysis in Schwarzschild hairy black hole,  somehow as an illustrative example, shall elaborate its application.}

This paper is organized as follows. In Sec. \ref{sec:Master equation and echo phase diagram}, we derive general master equations of hairy Schwarzschild black hole for perturbation field spins $s=0,1,2$, having the effective potential we scan such black hole parameter space and find the phase diagram to distinct single-peak and double-peak region. In Sec. \ref{sec:GD QNM spectra}, we present the hierarchical structure of the multi-families QNM spectrum in the frequency domain and provide a physical classification based on the QNM eigenfunction spacial configurations. In Sec. \ref{sec:excitation coefficient and energy fraction}, we detail our windowed time-domain analysis, showcasing the energy competition between QNM families and its source sensitivity. Finally, we summarize our findings and discuss their implications for future observations in Sec. \ref{sec: conclusions}. Additionally, we review the GD approach in Appendix~\ref{appendix:background}, and demonstrate the supplemented derivation for general master equation in Appendix~\ref{appendix:master equation}. Details of the numerical methods for solving the master equation in the frequency domain and time domain are presented in Appendices \ref{appendix:hyperboloidal_framework}--\ref{sec:appendix_methodology}.

\section{Perturbation equation and echo parameter space}
\label{sec:Master equation and echo phase diagram}
We start from analyzing the perturbation equations, especially the effective potential to give the echo parameter space of the hairy Schwarzschild black hole \cite{Ovalle:2020kpd}. The construction of this black hole from GD approach and its extended studies are briefly reviewed in Appendix \ref{appendix:background}. To  facilitate our study, we copy here the metric \eqref{eq:hairy sol 2a} of hairy Schwarzschild black hole
\begin{eqnarray}\label{eq:hairy sol 2}
    &&\mathrm{d}s^2=-A(r)\mathrm{d}t^2+\frac{\mathrm{d}r^2}{B(r)}+C(r)(\mathrm{d}\theta^2+\sin^2\theta\mathrm{d}\phi^2)\, ,\\
    &&\text{with}\quad   A(r)=B(r)=1-\frac{2M}{r}+\alpha\, \mathrm{e}^{-\frac{r}{M-l_{0}/2}}\equiv f(r) \quad \text{and}\quad   C(r)=r^2\, .\label{eq:function_fa}
\end{eqnarray}
{Here, $M$ represents the asymptotic mass of the hairy black hole, which is related with the seed Schwarzschild mass $\mathsf{M_s}$, via $M=\mathsf{M_s}+l_0/2$. The parameters $\{\alpha,l_0\}$ originate from the GD sector, where the positive $\alpha$ quantifies the deformation strength and $l_0$ corresponds to the charge associated with primary hair. More details can be seen from Appendix \ref{appendix:background}. Note that $l_0 \leq 2M$ should be satisfied to guarantee the asymptotic flatness of the hairy black hole \cite{Ovalle:2020kpd} } This black hole was found to admit an additional PS in certain parameter regimes \cite{Guo:2022ghl}, comparing against the single PS in Schwarzschild black hole. These multiple PSs bring in richer and distinctive optical appearances \cite{Meng:2023htc,Ribeiro:2025ohn} and interferometric signatures \cite{Wang:2025hzu}. The presence of additional PS has also found to closely associated with the echo signals of the probe scalar field \cite{Cavalcanti:2022cga,Yang:2024rms}. Moreover, the QNM and greybody factor for various perturbations of this black hole has been studied in~\cite{Yang:2022ifo}, in which the authors focus on the case with single peak in the perturbative potentials.

Here, we shall carefully explore the QNM of perturbation around the hairy Schwarzschild black hole \eqref{eq:hairy sol 2}, especially, when the perturbative potential has double peaks. The master equation for massless perturbations with different field spins can be incorporated into the Regge-Wheeler-like formula
\begin{eqnarray}\label{eq: master eq}
    \Big(-\frac{\partial^2}{\partial t^2}+\frac{\partial^2}{\partial r_{\star}^2}-V_{s}\Big)\Psi_{s,\ell,m}=0\, ,\quad \text{with}\quad V_{s}(r)=f(r)\Big[\frac{\ell(\ell+1)}{r^2}+(1-s)\frac{f^{\prime}(r)}{r}+\delta_{s,2}f^{\prime\prime}(r)\Big]\, ,
\end{eqnarray}
where $\mathrm{d}r_{\star}=\mathrm{d}r/f(r)$ denotes the tortoise coordinate, $\Psi_{s,\ell,m}$ denotes the master field reduced from dynamics variables of a certain perturbation equation. Whereas $s=0,1,2$ indicate scalar, electromagnetic and gravitational axial perturbations, respectively, and $\delta_{s,2}$ stands for the Kronecker delta.

To establish a direct correspondence with the properties of the  PS, our subsequent analysis of QNMs will concentrate on electromagnetic perturbations $(s=1)$. Although  the dynamics of all massless perturbations are known to converge to that of null geodesics \cite{Cardoso:2008bp}, as one can see that the master equations \eqref{eq: master eq} with different spins become identical in the eikonal limit $(\ell \gg 1)$, the QNMs of electromagnetic fields often provide a better numerical approximation to the geometric optics results for finite, low-lying multipoles.

Before proceeding, we shall briefly review the reducing procedure of the master scalar $\Psi_{1,\ell,m}$. For $s=1$, the perturbation could be sourced by an external electromagnetic field, of which the dynamics is depicted by the Maxwell equation
\begin{equation}
    \nabla_{\mu} F^{\mu\nu} = g^{\mu\alpha} \nabla_{\mu} (\partial_{\alpha}A^{\nu} - \partial_{\nu}A_{\alpha}) =0\, ,
\end{equation}
where the field strength tensor $F^{\mu\nu}$ is defined in terms of the vector potential $A_{\mu}$ as $F = \mathrm{d}A$. In a spherically symmetric background, the vector potential can be decomposed into a Regge-Wheeler-Zerilli (RWZ) form  \cite{Ruffini:1972pw,Chandrasekhar:1985kt}
\begin{eqnarray}\label{eq: EM decomposition}
    A_\mu=\sum_{\ell ,m}\left[
    \begin{pmatrix}
    0\\
    0\\
    A_{\ell, m}(t,r) \, \frac{1}{\sin\theta} \, \partial_\varphi Y_{\ell m}\\
    -A_{\ell, m}(t,r) \, \sin\theta \, \partial_\theta Y_{\ell m}
    \end{pmatrix}_{\text{odd}}
+\quad \begin{pmatrix}
        J_{\ell, m}(t,r) \, Y_{\ell, m}\\
        H_{\ell, m}(t,r) \, Y_{\ell, m}\\
        K_{\ell, m}(t,r) \, \partial_\theta Y_{\ell, m}\\
        K_{\ell, m}(t,r) \, \partial_\varphi Y_{\ell, m}
        \end{pmatrix}_{\text{even}}\right]\, ,
\end{eqnarray}
where for axial modes with odd parity $(-1)^{\ell+1}$, we can simply set
\begin{eqnarray}\label{eq:EM odd master scalar}
    A_{\ell,m}=\Psi^{\text{odd}}_{1,\ell,m}\, .
\end{eqnarray}
For polar modes with even parity $(-1)^{\ell}$, the master scalar can be taken as \cite{Chandrasekhar:1985kt}
\begin{eqnarray}\label{eq:EM even master scalar}
    \dot{K}_{\ell,m}&=&\sqrt{A\,B}\, \dot\Psi^{\text{even}}_{1,\ell,m}+J_{\ell,m}=f\, \dot\Psi^{\text{even}}_{1,\ell,m}+J_{\ell,m}\, ,\nonumber \\
    \dot{H}_{\ell,m}&=& \sqrt{\frac{A}{B}}\,\frac{\ell(\ell+1)}{C}(\Psi^{\text{even}}_{1,\ell,m})'+J_{\ell,m}'=\frac{\ell(\ell+1)}{r^2} (\Psi^{\text{even}}_{1,\ell,m})'+J_{\ell,m}'\, ,
\end{eqnarray}
where the dot represents the time derivative, and the prime represents the radial derivative. The  decomposition \eqref{eq: EM decomposition} divides the Maxwell equation into odd part and even part, and introducing of master scalar \eqref{eq:EM odd master scalar}-\eqref{eq:EM even master scalar}  then reduces the both equations into a same master equation \eqref{eq: master eq} with $\Psi^{\text{odd}}_{1,\ell,m}=\Psi^{\text{even}}_{1,\ell,m}=\Psi_{1,\ell,m}$.
Noted that the parallel procedures for other spins are shown in Appendix \ref{appendix:master equation}.

We move on to investigate the morphology of the effective potentials. A crucial aspect of the hairy black hole solution \eqref{eq:hairy sol 2} is the rich phenomenology that arises from its full parameter space. While the analysis in \cite{Cavalcanti:2022cga,Avalos:2023jeh} focused on a specific subspace by imposing the constraint $l_0=2M\alpha/(\mathrm{e}^2+\alpha)$, {with $\mathrm{e}$ the base of the natural logarithm,} which adjusts the horizon radius of the hairy black hole horizon radius to that of Schwarzschild black hole (e.g. $r_{h}(M,\alpha,l_0)=2\mathsf{M}_s$). Here we shall explore the broader implications by treating the hairy charge $l_0$ and the deformation intensity ${\alpha}$ as independent variables.  The interplay between these two parameters can induce a non-trivial structure in the spacetime geometry. Specifically, for certain regions in the $\{\alpha, l_0 \}$ parameter space, the metric function $f(r)$ develops two inflection points outside the horizon. This non-monotonicity of the gravitational potential is directly imprinted onto the perturbative dynamics, causing the effective potential $V_s(r)$ to exhibit three local extrema.

\begin{figure}[htbp]
    \centering
        \subfigure[]{
        \includegraphics[width=0.33\linewidth]{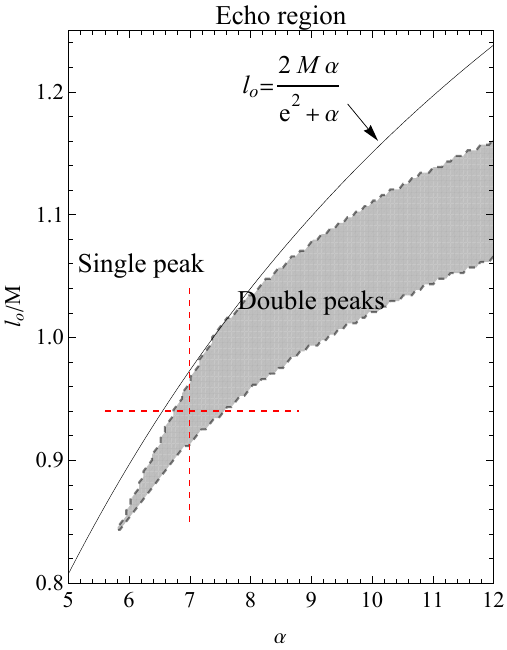}
        \label{fig:peak-number-phase-diagram}}
        \hfill
        \begin{minipage}[b]{0.61\textwidth}
        \centering
        \subfigure[]{
        \includegraphics[width=\linewidth]{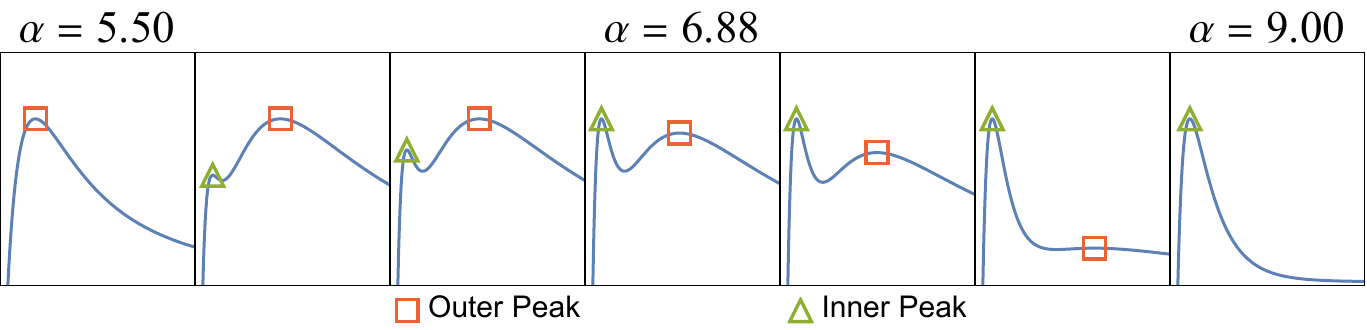}
        \label{fig:potential-horizontal-traversal}}
         \\[1.7cm]
        \subfigure[]{
        \includegraphics[width=\linewidth]{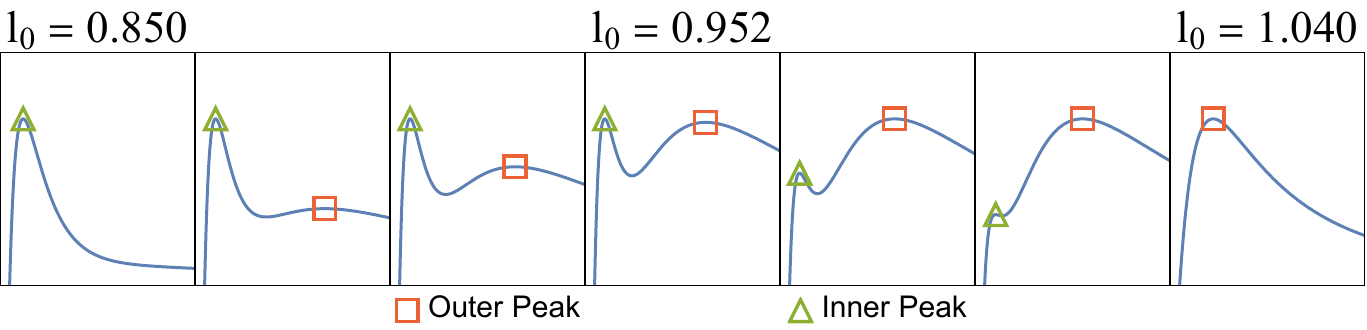}
        \label{fig:potential-vertical-traversal}}

         \end{minipage}
          \vspace{0.3cm}
	\caption{{\bf (a):} Parameter spaces of the effective potential morphology for electromagnetic perturbations with $s=1$, $\ell=6$ and $M=1$. The grey shaded region indicates the $\{\alpha, l_0\}$ parameter space where the potential exhibits a double-peak structure, a prerequisite for echoes. The blank region corresponds to a typical single-peak potential. The black curve shows the constraint $l_0=2M\alpha/(\mathrm{e}^2+\alpha)$ where the hairy black hole's horizon coincides with the seed Schwarzschild radius. The horizontal and vertical red dashed lines mark the potential varying along typical parameter paths analyzed in (b) and (c), respectively.
    {\bf (b):} Typical potential profiles along the horizontal path from (a) at a fixed $l_0/M=0.94$. This path demonstrates the transition from the Outer-Single-Peak regime (left end) to the Inner-Single-Peak regime (right end) as $\alpha$ increases.
    {\bf (c):} Typical potential profiles along the vertical path from (a) at a fixed $\alpha=7$. This path demonstrates the transition from the Inner-Single-Peak regime (bottom end) to the Outer-Single-Peak regime (top end) as $l_0$ increases.  {Note that in {\bf (b)} and {\bf (c)}, the {red squares} mark an outer single peak which survives from the outer barrier of the double-peak phase as the parameter changes, while the
     {green triangles} mark an inner singe peak surviving from the inner barrier.}
     }
\end{figure}

{We then scan the parameter $\{\alpha, l_0 \}$ and map the morphology of the electromagnetic potential with fixed  $s=1$, $\ell=6$ and $M=1$ (same settings for later studies) in  Fig. \ref{fig:peak-number-phase-diagram}, where the gray shaded region highlights the parameter combinations that give rise to the double-peak potential, in contrast to the blank region corresponding to the typical single-peak case. To further explore the phenomenological consequences of this transition, we must select specific paths through this parameter space for our subsequent QNM analysis. To this end, we choose two representative parameter trajectories that traverse the borders between the single-peak and double-peak regions, as indicated by the red dashed lines in Fig. \ref{fig:peak-number-phase-diagram}: one horizontal with fixed $l_0/M=0.94$  and one vertical with fixed $\alpha=7$. The behavior of effective potential is shown in Figs. \ref{fig:potential-horizontal-traversal} and \ref{fig:potential-vertical-traversal}, respectively.  In Fig. \ref{fig:potential-horizontal-traversal} with fixed $l_0/M=0.94$, we see that as $\alpha$ deceases, the inner peak marked by green triangle disappears while the outer peak marked by red Square survives, indicating the entering of Outer-Single-Peak regime. Inversely, when increasing $\alpha$, the outer peak disappears while the inner peak survives, indicating the entering of Inner-Single-Peak regime. Therefore, along the horizontal parameter trajectories with fixed $l_0/M=0.94$, the effective potential is in the Outer-Single-Peak regime, double-peak regime and Inner-Single-Peak regime as $\alpha$ increases. Similar phenomena is obvious for the vertical parameter trajectories with fixed $\alpha=7$, as shown in Fig. \ref{fig:potential-vertical-traversal}.
Then by analyzing the QNM spectra along these paths, we aim to uncover the distinctive phenomenological signatures associated with the emergence of the potential well, providing a direct link between the spacetime geometry and its observable ringdown properties. As we will see soon, this classification of effective potentials based on the number of peaks, especially further tracing the outer and inner peak, is important for disclosing the global structure of the QNM spectrum.}

\section{Photon sphere and echo QNM families}
\label{sec:GD QNM spectra}

Upon performing a temporal Fourier decomposition on the master scalar, $\Psi_{s,\ell,m}(t,r) = \mathrm{e}^{-\mathrm{i}\omega t} \Psi_{s,\ell,m}(r)$, the master equation \eqref{eq: master eq} transforms into a one-dimensional Schrödinger-like equation for the complex frequency $\omega$, namely
\begin{eqnarray}\label{master_equation_frequency_domian}
    \mathcal{M}[\Psi]\equiv\left[\frac{\mathrm{d}^2}{\mathrm{d}r_{\star}^2}+\omega^{2}-f(r)\left(\frac{\ell(\ell+1)}{r^2}+(1-s)\frac{f^{\prime}(r)}{r}+\delta_{s,2}f^{\prime\prime}(r)\right)\right]\Psi_{s,\ell,m}(r)=0\, .
\end{eqnarray}
The dissipative nature of the black hole spacetime imposes specific boundary conditions on the radial function $\Psi_{s,\ell,m}(r)$. Classically, a  QNM is defined as a solution that is purely ingoing at the event horizon ($r_{\star} \to -\infty$) and purely outgoing at spatial infinity ($r_{\star} \to \infty$) \cite{Nollert:1999ji,Berti:2009kk,Konoplya:2011qq}. In our present study, the effective potential $V_s$ vanishes both at the event horizon and  spatial infinity. This translates to the asymptotic behavior of $\Psi_{s,\ell,m}(r)$, i.e.,
\begin{equation}\label{boundary_conditions_QNMs}
    \Psi_{s,\ell,m}(r_{\star}\rightarrow -\infty)\sim \mathrm{e}^{-\mathrm{i}\omega r_{\star}}, \quad \text{and} \quad \Psi_{s,\ell,m}(r_{\star}\rightarrow \infty)\sim \mathrm{e}^{\mathrm{i}\omega r_{\star}}.
\end{equation}

Solving for QNMs involves dealing with the boundary conditions and the divergence of eigenfunctions at the boundaries. The hyperboloidal framework offers a powerful approach to overcome these difficulties~\cite{Zenginoglu:2007jw,Ansorg:2016ztf,PanossoMacedo:2018hab,Zenginoglu:2011jz, PanossoMacedo:2023qzp, Zenginoglu:2024bzs, PanossoMacedo:2024nkw,Jaramillo:2020tuu}. By using constant-$\tau$ slices that extend to null infinity and penetrate the event horizon, the hyperboloidal framework prevents the divergence of eigenfunctions. Moreover, at the boundaries, light cones are oriented outward from the computational domain, so the boundary conditions are naturally satisfied. The divergent boundary conditions (\ref{boundary_conditions_QNMs}) are replaced by a requirement of global regularity for the eigenfunction across the compactified hyperboloidal slices.

The QNM spectra we obtain by using the pseudospectral method, which is well-suited to the hyperboloidal framework, reveals a remarkably rich structure, particularly in the parameter space near the echo-generating region. This finding motivates us to carry out a more thorough investigation, employing a range of complementary computational techniques to cross-validate the results and to build a complete physical picture.
As we will see soon that the double-peak potential in the hairy Schwarzschild black hole leads to a clear bifurcation of the spectrum. Aided by the WKB approximation as a classification tool, we distinct them as PS family and echo family, respectively, meaning that the deformation source not only shift the QNM spectra but also fundamentally alters the organization and classification of the spectra families. Thus,
instead of the individual QNMs and their collection in traditional study, here we shall focus more on the global structure of the perturbation spectrum.

\subsection{Methodology for QNM calculation and validation}
\label{sec:Methodology for QNM Calculation}

To obtain a reliable picture of the QNM spectra, we employ a suite of distinct methods, each with its own advantage, allowing for cross-validation. We briefly list them as follow and more details are elaborated in the appendices.

\begin{enumerate}
    \item \textbf{Pseudospectral Method (Primary Method):} Our results are primarily obtained using the pseudospectral method, which transforms the differential operator eigenvalue problem into a matrix eigenvalue problem with the Chebyshev-Lobatto grid being used. The hyperbolidal framework of this hairy black hole is presented in Appendix \ref{appendix:hyperboloidal_framework}.

   \item \textbf{WKB Approximation (for Classification):} The WKB method, pioneered by Schutz and Will \cite{Schutz:1985km}, provides a semi-analytic tool. Its key idea is to connect the solution near a potential extremum with the asymptotic wave solutions. For a potential barrier with $V^{\prime\prime}(r_0)<0$ at the peak $r_0$, matching solutions across the peak yields the well-known quantization condition for complex QNM spectra
  \begin{eqnarray} \label{eq:wkb-formula}
        \frac{Q_0}{\sqrt{2Q_0''}} = \mathrm{i}\Big(n+\frac{1}{2}\Big)\, ,
    \end{eqnarray}
    where $Q(r_{\star}) \equiv \omega^2 - V(r_{\star})$, and $Q_0$, $Q_0''$ are evaluated at the peak.  Since $Q_0'' = -V_0'' > 0$, this formula naturally yields a complex $\omega$. More details can be referred to \cite{Konoplya:2003ii}, however, we shall give some comments on its application on a potential well with $V''(r_0)>0$ at the well $r_0$. Since the term $Q_0'' = -V_0''$ then becomes negative, the square root $\sqrt{2Q_0''}$ thus produces an imaginary factor, which cancels the $i$ on the right-hand side, leading to a condition for a purely real frequency. Furthermore, the asymptotic WKB solutions far from the turning points take the form,
    \begin{equation}
        \Psi_{\text{WKB}}(r_{\star}) \approx Q(r_{\star})^{-1/4} \exp\Big(\pm \mathrm{i} \int \sqrt{Q(t)} \mathrm{d}t\Big).
    \end{equation}
    For a potential well, a bound state requires $\omega^2 < V(r_{\star} \to \pm\infty)$, causing $Q(r_{\star})$ to become negative in the asymptotic regions. Consequently, $\sqrt{Q}$ becomes imaginary, and the oscillatory solution transforms into an exponentially decaying one, characteristic of a bound state. This dual reasoning explains why the WKB approximation interprets modes trapped in a potential well as purely real bound states.

    \item \textbf{Direct Integration Method (Shooting Method):} In this approach, the master equation (\ref{eq: master eq}) is integrated from the boundaries, and the QNM spectra are determined by requiring a smooth match of the solutions in an intermediate region. It is known to be very robust for solving quasi-bound states (echo modes), and its reliability for computing bound states has also been highlighted recently~\cite{Volkel:2025lhe}. Here we mainly employ this method to validate the imaginary part of the echo modes, thus offering a crucial verification of their damping rates. For a detailed account of our implementation, see Appendix \ref{sec:appendix_direct integration}.

    \item \textbf{Time-Domain Analysis with Matrix Pencil:} We perform time-domain evolutions of the master equation and extracted the QNM spectra from the late-time ringdown signal using the Matrix Pencil (MP) method~\cite{Berti:2007dg,56027,370583}. This technique is known for its high accuracy and provides a completely independent validation route. A brief overview is given in Appendix~\ref{sec:appendix_methodology}.
\end{enumerate}

\subsection{Analysis of the QNM spectrum and method comparison}
Using the methods listed above, we calculate the QNM spectra for selected parameter sets with fixed $\alpha=7$, showing in Tab. \ref{table:qnm-comparison},
which provides a detailed look at the QNM spectra for three representative cases and highlights the agreement between different computational methods. More importantly, it reveals the structural changes in the spectra as the effective potential transiting between single-peak and double-peak configurations. {Specifically, the discrepancy between WKB predictions and pseudospectral results manifests differently depending on the potential's profile. The pseudospectral method, validated by the matrix pencil method, proves essential for identifying the full spectrum where the WKB approximation serves as a preliminary mode classifier. We discuss these specific manifestations below: }

\begin{table}[ht!]
    \centering
\begin{tabular}{|c|c|c|c|c|}
    \hline
    & \textbf{Overtone (n)} & \textbf{Pseudospectral} & \textbf{Time-Domain (MP)} & \textbf{9th-WKB} \\
    \hline
    \multirow{4}{*}{\shortstack{Inner-Single-Peak \\ $l_0/M=0.890$}}
                               & 0 & 1.444794-0.074881 i& 1.444794-0.074881 i&  -  \\\cline{2-5}
                               & 1 & 1.554661-0.167495 i& 1.554658-0.167496 i&  -  \\\cline{2-5}
                               & 2 & 1.679244-0.264020 i& 1.679484-0.264193 i&  -  \\\cline{2-5}
                               & 3 & 3.289470-0.286500 i& 3.289550-0.286691 i& 3.289452-0.286485 i\\\hline
        \multirow{4}{*}{\shortstack{Double-Peak \\ $l_0/M=0.952$}}
                               & 0 & 1.121228-1.56923$\times 10^{-7}$ i& 1.121228-1.56862$\times 10^{-7}$ i& 1.121481 \\\cline{2-5}
                               & 1 & 1.195112-1.35497$\times 10^{-5}$ i& 1.195112-1.35498$\times 10^{-5}$ i& 1.196931 \\\cline{2-5}
                               & 2 & 1.261775-4.34136$\times 10^{-4}$ i& 1.261776-4.34135$\times 10^{-4}$ i& 1.268977 \\\cline{2-5}
                               & 3 & 1.318228-4.87563$\times 10^{-3}$ i& 1.318229-4.87563$\times 10^{-3}$ i& 1.338180 \\\hline
        \multirow{4}{*}{\shortstack{Outer-Single-Peak \\ $l_0/M=1.040$}}
& 0 & 1.276954-0.079749 i& 1.276954-0.079749 i& 1.276953-0.079748 i\\\cline{2-5}
                               & 1 & 1.270543-0.239527 i& 1.270539-0.239520 i& 1.270537-0.239525 i\\\cline{2-5}
                               & 2 & 1.257793-0.400135 i& 1.256677-0.399826 i& 1.257770-0.400132 i\\\cline{2-5}
                               & 3 & 1.238840-0.562103 i& 1.216388-0.571014 i& 1.238783-0.562110 i\\\hline
    \end{tabular}
    \caption{Comparison of electromagnetic QNM spectra $(M \omega)$ for selected parameter sets with $\alpha=7$. The results in the table show the  agreement for the QNM spectra between pseudospectral method and various validation techniques. The symbol `-' indicates the WKB formula is blind for those overtones, while a purely real value indicates a bound-state prediction for a potential well.}
    \label{table:qnm-comparison}
\end{table}

\begin{enumerate}

\item For the \textbf{Inner-Single-Peak case} $(l_0/M=0.890)$, the fundamental mode predicted by the WKB method does not correspond to the fundamental mode $(n=0)$ of the pseudospectral results. Instead, it aligns with a higher overtone $(n=3)$. This noteworthy feature indicates that the presence of a new mode family dominates the low-damping part of the spectrum. This feature is robustly confirmed by the Time-Domain (MP) method, ruling out the possibility that these are spurious modes of the pseudospectral method.

\item For the \textbf{Outer-Single-Peak case} ($l_0/M=1.040$), the WKB fundamental mode matches the pseudospectral fundamental mode $(n=0)$. This scenario represents the familiar behavior expected from black holes with a typical potential with a single peak, where the spectrum is dominated by a single family of modes associated with the main potential barrier.

\item The \textbf{double-peak case} ($l_0/M=0.952$) showcases the most interesting phenomenology. These modes, characterized by their extremely low decay rate and a distinct echo period of $\mathcal{T}=\frac{2\pi}{\Delta \text{Re}(\omega_n)}$, are identified by both the Time-Domain (MP) and pseudospectral methods, with the results showing agreement. Furthermore, we employ the WKB approximation around the potential well, which gives a series of purely real, bound-state-like modes. They show good agreement with the real parts of the extremely long-lived echo modes found by other two methods.
\end{enumerate}
The above analysis hints the existence of a new family of QNMs and highlights the efficacy of the WKB method as a mode classifier. This motivates us to scan the parameter space to specify the classification of the QNM spectrum.

\subsection{Classification and evolution of QNM families}
\begin{figure}[htbp]
	\centering
	\includegraphics[width=0.44\textwidth]{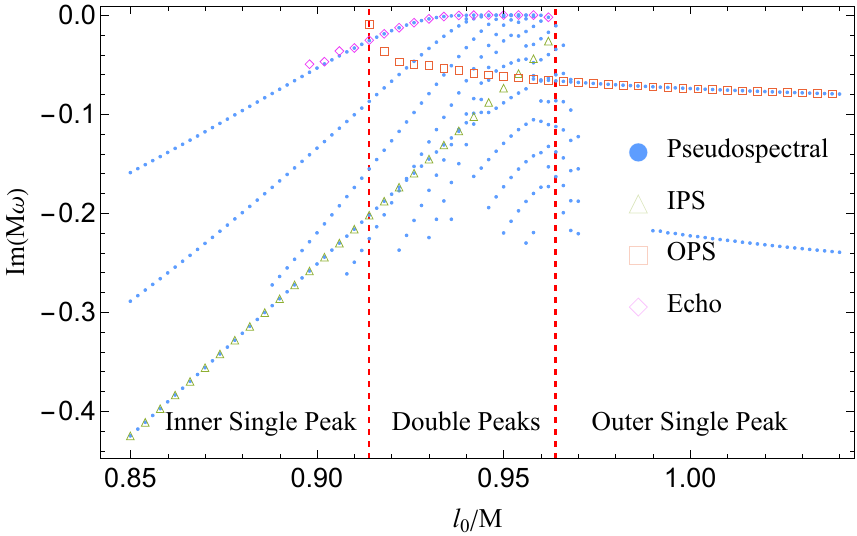}\hspace{1cm}
	\includegraphics[width=0.42\textwidth]{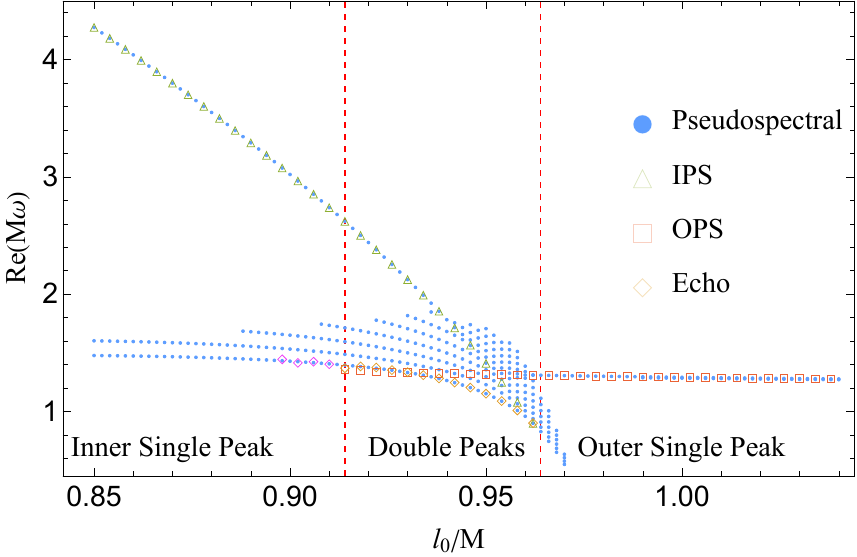}\\
        \includegraphics[width=0.44\textwidth]{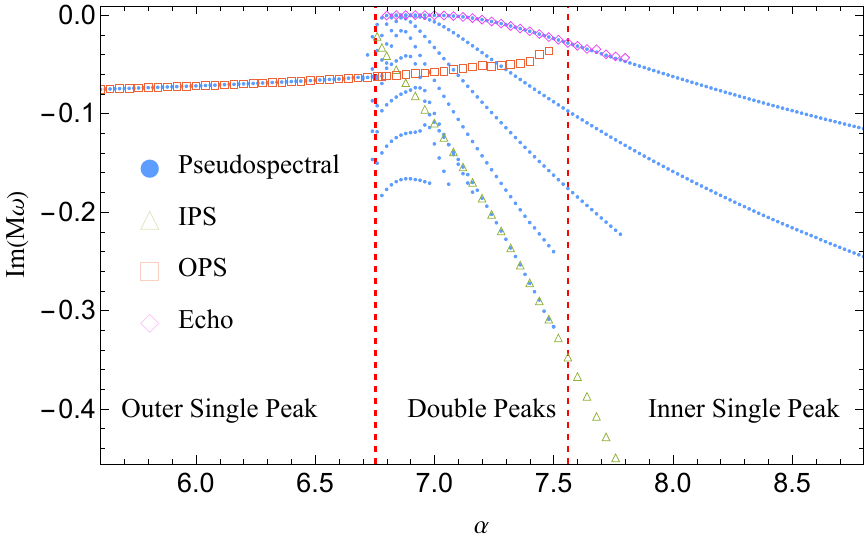}\hspace{1cm}
	\includegraphics[width=0.43\textwidth]{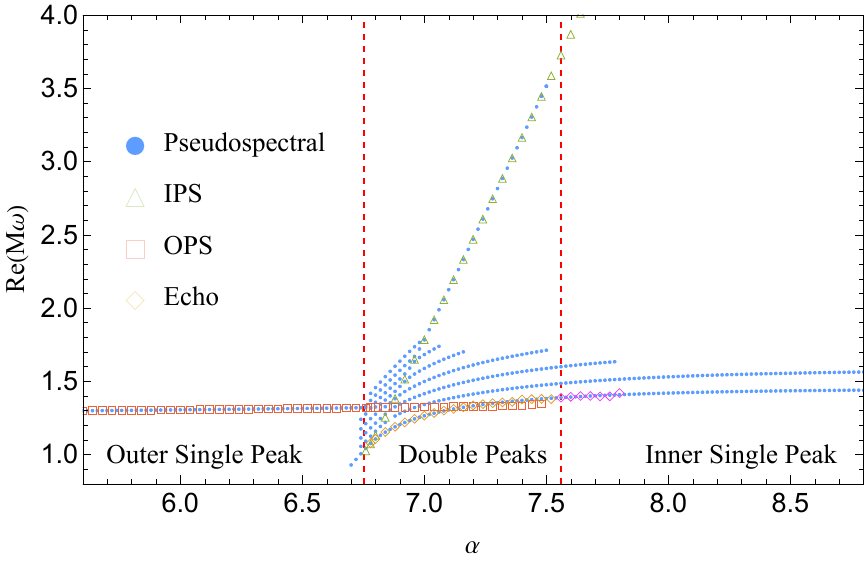}
	\caption{\textbf{Top Panel}: The imaginary (left) and real (right) parts of the QNM spectra as  $l_0/M$ is varied at  fixed  $\alpha=7$. This corresponds to a vertical traversal of the parameter space shown in Fig. \ref{fig:peak-number-phase-diagram}. \textbf{Bottom Panel}: The imaginary (left) and real (right)   parts of the QNM spectra as $\alpha$ is varied at a fixed  $l_0/M=0.94$. This corresponds to a horizontal traversal. In each panel, the vertical red dashed lines mark the boundaries between the single-peak and double-peak (echo) regions. \textbf{Blue Dots}: High-precision data from the pseudospectral method, filtered by a spectral convergence test (drift ratio $\ge 10^3$~\cite{Chen:2024mon,BOYD199611}). \textbf{Colored Polygons}: The squares, triangles, and other polygons represent the first mode estimated by WKB approximation obtained from different extrema of the effective potential. They serve to classify the pseudospectral method  results into: the \textit{PS family} (OPS mode for the Outer-Single-Peak, IPS mode for the Inner-Single-Peak) and the \textit{echo family} (associated with the potential well). \textbf{Purple Diamonds}: These points are calculated using the direct integration method. They provide an independent validation for the imaginary part of the long-lived echo modes, where the WKB approximation is not applicable. The excellent agreement confirms the accuracy of our pseudospectral method results for the damping rates of these modes.}
	\label{fig:QNMs-curve}
\end{figure}

We scan the parameters along two representative trajectories that traverse the echo parameter space defined by the red dashed lines in Fig. \ref{fig:peak-number-phase-diagram}. The results of QNM spectra are presented in Fig. \ref{fig:QNMs-curve}, where the top panel corresponds to the vertical traversal with fixed $\alpha=7$ while the bottom panel is for horizontal traversal with  fixed $l_0/M=0.94$. It is worth emphasizing that the QNM spectra obtained from each method are continuous functions of the black hole parameters. Nevertheless, in  Fig. \ref{fig:QNMs-curve}, we represent the QNM frequencies with distinct  discrete markers, solely to facilitate their visual distinction and classification.

The WKB approximation serves as a guide to the physical origin of the different mode families. %{However, as we will see soon, a more intrinsic and generic classification approach, one that is model-independent, lies in analyzing the spatial profiles of the QNM eigenfunctions. In particular, having the hairy Schwarzschild black hole serve as an illustrative example, the distinct physical nature of different QNM families is clearly revealed by the varying degrees of localization in their eigenfunction configurations.}
As seen in the top panels of Fig. \ref{fig:QNMs-curve}, in the Outer-Single-Peak region, the WKB approximations for the Outer Photon Sphere (OPS) modes (red squares) obtained from the outer peak show excellent agreement with the pseudospectral method results (blue dots). This represents the familiar GR-like scenario. However, upon entering the center of double-peak region, the WKB markers for both the OPS and Inner Photon Sphere (IPS) modes obtained from the inner peak lose their accuracies. This failure is expected, as the WKB approximation breaks down when the potential develops a complex, multi-extremum structure.

\begin{figure}[htbp]
	\centering
	 \includegraphics[width=0.275\textwidth]{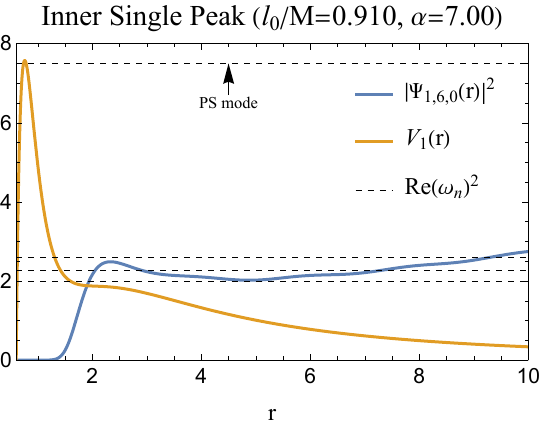}\hspace{0.2cm}
	 \includegraphics[width=0.282\textwidth]{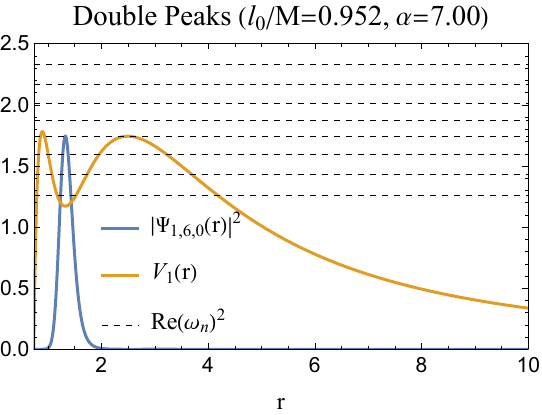}\hspace{0.2cm}
  \includegraphics[width=0.28\textwidth]{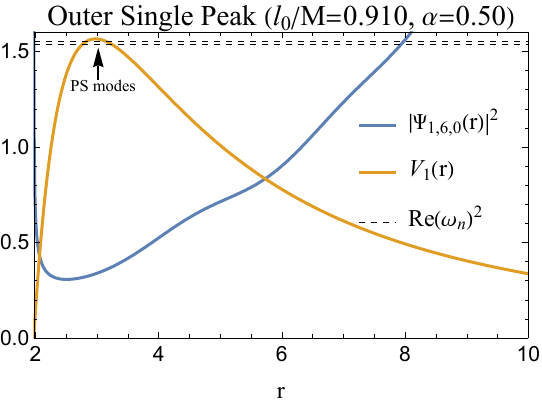}
	\caption{
    \textbf{Left Panel:} In the Inner-Single-Peak region, the spacial configuration of the fundamental mode (with $M\omega_0=1.405711-0.033302~ \mathrm{i}$)  is significantly condensed near the potential's ``shoulder'', demonstrating the diffraction trapping effect. \textbf{Middle Panel:} In the double-peak region, the spacial configuration of the fundamental mode (with $M\omega_0=1.121228-1.569230\times 10^{-7} ~ \mathrm{i}$) is highly localized within the potential well, exhibiting a clear quasi-bound state (QBS) character. \textbf{Right Panel:} In the Outer-Single-Peak region, the spacial configuration of the fundamental mode (with $M\omega_0=1.246082-0.096051~ \mathrm{i}$)  is highly delocalized, which is a characteristic of a standard scattering resonance associated with the PS. In each panel, the black dashed lines indicate the energy level, $\text{Re}(\omega_n)^2$, of the corresponding QNM from the pseudospectral data.}
	\label{fig:eigenfunctions}
\end{figure}

An interesting phenomenon occurs during the transition from the double peaks to the Inner-Single-Peak region. Unlike the abrupt disappearance of echo modes when transitioning to the Outer-Single-Peak region, the echo family extends smoothly into the Inner-Single-Peak domain. By a quick glance on Fig. \ref{fig:potential-vertical-traversal}, we argue that this is because the disappearance of the outer peak leaves behind a prominent ``shoulder" in the effective potential. This shoulder creates a region where the potential decays slowly towards spatial infinity, leading to a diffraction trapping effect for waves. {Noted that  the existence of such modes has also been independently found in a recent work \cite{Wang:2025mxe} in which the authors  identified them as the off-peak modes.} This underlying physical mechanism can be intuitively understand when one focus on the spacial configuration of a single mode, as it can be explicitly seen in Fig. \ref{fig:eigenfunctions}, where we present the fundamental eigenfunction ($\Psi_{s,\ell,m}=\Psi_{1,6,0}$, blue curves) , computed by the direction integration method, against their related  effective potentials ($V_s=V_1$, yellow curves) for each case.

In the Outer-Single-Peak case (right panel of Fig. \ref{fig:eigenfunctions}), which resembles the standard GR scenario, the eigenfunction is highly delocalized, spreading over a wide spatial range and its energy level $\text{Re}(\omega_0)^2$ approaches the peak of potential barrier. This is the hallmark of a typical scattering resonance, where the wave interacts briefly with the potential barrier before efficiently dissipating to the boundaries, corresponding to a short-lived mode with a large imaginary part.

In contrast, the double-peak case (middle panel of Fig~\ref{fig:eigenfunctions}) features a highly localized eigenfunction, whose energy level, $\text{Re}(\omega_0)^2$, lies above the bottom of the potential well. As a result, its amplitude almost entirely confined to this region. This is a clear signature of a quasi-bound state (QBS), where the wave is effectively trapped, leading to a long-lived echo mode with an extremely small imaginary part.

The Inner-Single-Peak case (left panel of Fig. \ref{fig:eigenfunctions}) represents a fascinating intermediate state. The eigenfunction, of which the energy level $\text{Re}(\omega_0)^2$ approaches the potential's  ``shoulder", is partially localized, showing a significant condensation of its amplitude near this ``shoulder." This configuration perfectly illustrates the diffraction trapping mechanism, a mode that is more confined than a simple scattering resonance but more dissipative than a typical QBS. Thus, analyzing the eigenfunction's degree of localization provides a potential method for classifying the QNM families.

This re-classification also explains the reordering of the spectral hierarchy observed in Fig. \ref{fig:QNMs-curve}. In the Inner-Single-Peak region, the partially trapped echo family becomes the set of fundamental modes. Consequently, the original IPS modes, being highly delocalized scattering resonances, are relegated to highly damped, excited states. The fate of these IPS modes exhibits a subtle dependence on the path taken through the parameter space. Along the vertical traversal (top panel of Fig. \ref{fig:QNMs-curve}), the WKB approximation (green triangles) can still track the IPS modes as they extend continuously from the double-peak region into the Inner-Single-Peak region. However, along the horizontal traversal (bottom panel of Fig. \ref{fig:QNMs-curve}), the WKB approximation for the IPS modes vanishes abruptly at the transition boundary. This occurs because as the parameters move deeper into the Inner-Single-Peak region along this path, the damping rate of these IPS modes increases dramatically, making them numerically challenging to resolve and requiring higher grid resolution to maintain their convergences above our threshold.

\section{Complex amplitude and energy fraction}
\label{sec:excitation coefficient and energy fraction}
The analysis in the previous section revealed a rich structure, particularly the competition between the PS and echo QNM families. Then natural questions arise: when two or more QNM families coexist, which one is the dominant mode and how to determine their dominances?  An usual treatment is to consider the fundamental  modes with the smallest absolute of imaginary part as the dominant mode, but these modes may only become dominant at very late times, and a waveform reconstructed from it would miss crucial information from the early-time signal. Thus, it was addressed in~\cite{Giesler:2019uxc} that including overtones allows for the modeling of the black hole ringdown signal for all times, stressing the importance of the overtone. On the other hand, realistic observation of long-live ringdown signals may  inevitably be ``windowed'', as such mode's characteristic lifetime can be extraordinarily long.  For instance, using the fundamental echo mode data from Tab. \ref{table:qnm-comparison}, their lifetimes for a supermassive hairy Schwarzschild black hole with $M \sim 10^7 M_\odot$ would be approximately $\tau=\frac{1}{|\text{Im}(M\omega)|}\frac{GM}{c^3}\approx 9.9$  years, which can be comparable to the entire mission duration of space-based detectors.

Thus, it is significant to develop  the windowed analysis of the ringdown waveform. To start with, let us briefly review the standard procedure of fitting ringdown waveform in a data-driven framework. We shall firstly solve the  master equation \eqref{eq: master eq} using the finite difference method (see Appendix \ref{sec:appendix_finite difference} for details) to obtain the time-domain profile of the master scalar $\Psi(t)$. Second, we model the resulting time-series data based on the linear superposition of damped exponents ansatz, $\Psi(t)\simeq \sum_{k=1}^{p} h_{k}\,\mathrm{e}^{-\mathrm{i}\omega_k t}$,  where the sum over $p$ includes all relevant modes $\{\omega_{n}\}$ and their mirror modes $\{-\overline{\omega_{n}}\}$. Assuming that the modeling ringing phase starts at the peak of the signal $t=t_{\text{peak}} = 0$ and ends at $t=N\,\Delta t_{\text{eff}}$, where $ \Delta t_{\text{eff}}$ is the sampling time step and $N\ge2p-1$ is an integer, we can write the discrete time-series signal $\Psi(m\, \Delta t_{\text{eff}}) \equiv x[m]$ as
\begin{equation}
    x[m] = \sum_{k=1}^{p} h_k (z_k)^{m}\, , \quad \text{with} \quad z_k = \mathrm{e}^{-\mathrm{i}\omega_k \Delta t_{\text{eff}}}\, ,
    \label{eq:signal_model}
\end{equation}
where $h_k$ is the complex amplitude of the $k$-th mode for the master scalar $\Psi(t)$. To connect this amplitude to a physical observable, we note that the outgoing EM radiation at infinity is described by the Maxwell-Newman-Penrose scalar $\mathbf{\phi}_{2}$ \cite{Teukolsky:1973ha}, and as demonstrated by Chandrasekhar \cite{Chandrasekhar:1985kt}, the RWZ master scalar $\Psi$ and the $\mathbf{\phi}_{2}$ are linked by a well-established transformation involving a first-order differential operator. For any given mode $k$ oscillating as $\mathrm{e}^{-\mathrm{i}\omega_k  t}$, this relationship manifests in the frequency domain as a direct proportionality to the mode's frequency. Therefore, the complex amplitude $h_{k}$ of the master scalar can be related to the QNM excitation coefficient $C_{k}$ \cite{Berti:2025hly}. As addressed in early foundational works \cite{Andersson:1996cm,Leaver:1986gd,Berti:2006wq,Oshita:2021iyn}, the QNM excitation coefficient $C_{k}=E_{k}T_{k}$ can be further factorized into a source-independent excitation factor $E_{k}$ and a source-dependent term $T_{k}$. The factor $E_{k}$ is determined by the residue of the Green's function at the corresponding  pole, while $T_{k}$ depends on the initial data and on the specific source exciting the perturbations \cite{Andersson:1996cm}. This inspired more studies on the QNM  excitation coefficient, see for examples \cite{Oshita:2025ibu,Oshita:2024wgt,Lo:2025njp}
and references therein. More recently, a definitive, cross-validated dataset for the Kerr excitation factors has been established \cite{Motohashi:2024fwt,Kubota:2025hjk}, in which the authors claimed to resolve ambiguities present in the earlier literature. Here we focus on the complex amplitude $h_k$ for our following purpose. As suggested in \cite{Berti:2007dg}, one can  robustly extract $\{\omega_{k}\}$ via applying the MP method  and subsequently  determine $\{h_{k}\}$ using a linear least-squares fit, the main steps of which are present in Appendix \ref{sec:appendix_Matrix pencil} and Appendix \ref{sec:appendix_least-squares}, respectively.

\subsection{Defining energy fraction in a sliding observation window}
Since the observation window is always finite, so beside the  complex amplitude $h_k$ or excitation coefficient $C_{k}$, the decay of a mode is also important for the received ringdown signal. The latter is significant if the system has long-lived modes, especially for the echo mode family found in the current hairy Schwarzschild black hole. Thus, in order to combine both effects in determining the dominant mode, we shall define the new energy fraction to quantify the contribution of each mode within a finite window.

To proceed, we consider a finite observation window {starts at} $t_{0}=d\,\Delta t_{\text{eff}}$ {and ends at} $t_{\text{end}}=t_{0}+T=(d+\mathcal{N})\,\Delta t_{\text{eff}}$,  and then propose a time-shifting transformation
\begin{eqnarray}\label{eq:time_shifting_transformation}
    t \rightarrow t'&=&t- t_{0}\, ,\\
    m \rightarrow m'&=&m-d\, .
\end{eqnarray}
{Thus, such window profile for the shifted time coordinate becomes $t'\in [0,T]$.} Subsequently, the original signal \eqref{eq:signal_model} at indices $m'=0,1,2,\cdots,\mathcal{N}$ can be rewritten as
\begin{eqnarray}
    x[m'+d] = \sum_{k=1}^{p} h_{k} (z_k)^{(m'+d)} = \sum_{k=1}^{p} (h_{k} z_k^d) (z_k)^{m'}\equiv x_{\text{obs}}[m']=\sum_{k=1}^{p} \tilde{h}_{k}\, (z_k)^{m'}\, .
\end{eqnarray}
Here $\tilde{h}_{k}$ is indicated as an effective amplitude measured by the observation signal $x_{\text{obs}}[m']$ at $m'=0$, which is the primary observable quantity accessible to us. Especially, when the observation window is large enough, we have the relation
\begin{eqnarray}
    \tilde{h}_k(t_0) \equiv h_{k} (z_k)^d = h_{k} \mathrm{e}^{-\mathrm{i}\omega_k t_0}\, ,
    \label{eq:effective_h}
\end{eqnarray}
and its magnitude satisfies $|\tilde{h}_k| = |h_{k}| \mathrm{e}^{-|\text{Im}(\omega_k)|t_0}$, which  encodes the decay of the mode up to the start of our observation. Therefore, the energy radiated by a single QNM in an observation window  which is proportional to the time-integral of the square of the field's time derivative~\cite{Berti:2006hb}, can be reduced as
\begin{eqnarray}
    E_k(t_{0},T) &\propto& \int_{t_0}^{T+t_{0}} \left| \frac{\mathrm{d}\Psi_k(t)}{\mathrm{d}t} \right|^2 \mathrm{d}t \nonumber=\int_{t_0}^{T+t_{0}} |h_{k} (-\mathrm{i}\omega_k) \mathrm{e}^{-\mathrm{i}\omega_k t}|^2\mathrm{d}t \nonumber \\
    &=&\int_0^T |\tilde{h}_{k}(t_0) (-\mathrm{i}\omega_k) \mathrm{e}^{-\mathrm{i}\omega_k t'}|^2 \mathrm{d}t'\nonumber
    =|\tilde{h}_{k}(t_0)|^2 |\omega_k|^2 \int_0^T \mathrm{e}^{-2|\text{Im}(\omega_k)|t'} \mathrm{d}t' \nonumber \\
    &=&|\tilde{h}_{k}(t_0)|^2 |\text{Re}(\omega_k)|^2 \left( 1 + \frac{1}{4Q_k^2} \right) \int_0^T \mathrm{e}^{-2|\text{Im}(\omega_k)|t'} \mathrm{d}t'
    \approx |\tilde{h}_{k}(t_0)|^2 |\text{Re}(\omega_k)|^2 \int_0^T \mathrm{e}^{-2|\text{Im}(\omega_k)|t'} \mathrm{d}t', \label{eq:energy_approx}
\end{eqnarray}
where in the second line we have used  \eqref{eq:time_shifting_transformation} and \eqref{eq:effective_h} , while in the third line we have used the exact relation $|\omega_k|^2 = |\text{Re}(\omega_k)|^2(1 + 1/(4Q_k^2))$, where $Q_k = |\text{Re}(\omega_k)|/(2|\text{Im}(\omega_k)|)$ is the quality factor. For the QNM modes relevant to the current study, this quality factor is large. For instance, even for a relatively high overtone such as $n=8$ in our echo data, we find $Q_k^2 \approx 92$ and the echo data from~\cite{Guo:2022umh} is $Q_k^2 \sim 10^8$. Thus, the correction term $1/(4Q_k^2)$ is of the order of $10^{-3}$ or even smaller, such that the approximation in the last step can be safely employed. Further performing the final integral in \eqref{eq:energy_approx} shall give us the energy radiated by a single mode within the window
\begin{eqnarray}
    E_k(t_{0},T) \propto \frac{|\tilde{h}_{k}(t_0)|^2 |\text{Re}(\omega_k)|^2}{2|\text{Im}(\omega_k)|} \Big( 1 - \mathrm{e}^{-2|\text{Im}(\omega_k)|T} \Big)\, .
    \label{eq:energy_mode_windowed_final}
\end{eqnarray}
It is noted that in the idealized limits where the observation starts at the physical onset of the ringing $(t_0 \to 0)$ and the window is long enough to capture the entire signal $(T \to \infty)$, our expression \eqref{eq:energy_mode_windowed_final} reduces to the standard formula for the total radiated energy for a single multipole, $E_{\text{rad}} \propto |h_k|^2 |\text{Re}(\omega_k)|^2 / |\text{Im}(\omega_k)|$~\cite{Berti:2006hb}.

Sbusequently, we define the \textbf{Windowed Ringing Energy Fraction}, $\mathcal{E}_k(t_{0},T)$, for each mode as the ratio of its energy to the total energy summed over all contributing modes within the window,
\begin{equation}
    \mathcal{E}_k(t_{0},T) \equiv \frac{E_k(t_{0},T)}{\sum_{j} E_j(t_{0},T)} = \frac{ |\tilde{h}_{k}(t_0)|^2 |\text{Re}(\omega_k)|^2 |\text{Im}(\omega_k)|^{-1}( 1 - \mathrm{e}^{-2|\text{Im}(\omega_k)|T}) }{ \sum_{j} |\tilde{h}_{j}(t_0)|^2 |\text{Re}(\omega_j)|^2 |\text{Im}(\omega_j)|^{-1} ( 1 - \mathrm{e}^{-2|\text{Im}(\omega_j)|T}) }\, ,
    \label{eq:energy_fraction_windowed_final}
\end{equation}
which measures the normalized contribution of each QNM to the ringdown signal observed within the specific ringing phase duration $T$, allowing for an effective comparison of their excitation strengths under the ringdown simulation.

To close this subsection, we summarize our windowed analysis workflow of the time-domain waveform as
\begin{align}
\text{Master Eq. \eqref{eq: master eq}} \xrightarrow{\text{Finite Difference}} \Psi(t) \xrightarrow{\text{Matrix Pencil}} \{\tilde{\omega}_{k}\} \xrightarrow{\text{Least Squares}} \{\tilde{h}_{k}\} \xrightarrow{\text{Eq.~\eqref{eq:energy_fraction_windowed_final}}} \{\mathcal{E}_{k}(t_0)\},
\end{align}
along which we shall quantify the competition between QNM families within finite time windows in the next subsection.

\subsection{Competition for dominant modes}
To provide a comprehensive characterization of the competition between the echo and PS QNM families as the time goes by, we move on to analyze three distinct scenarios corresponding to different typical profiles of the double-peak potential.

\begin{enumerate}
    \item \textbf{Outer-Peak Dominant Case:}
    The profile of effective potential in this scenario  corresponds to the second panel of Fig. \ref{fig:potential-horizontal-traversal} as well as the inset of Fig. \ref{fig:Outer-Peak-Dominant-waveform} in tortoise coordinates. A typical time domain waveform is shown in Fig. \ref{fig:Outer-Peak-Dominant-waveform} with $\alpha=6.78$ and $\ell_0/M=0.940$, based on which we depict the  energy fraction in Fig. \ref{fig:Outer-Peak-energy-fraction1}. Also, the QNMs and coefficients extracted at $t_0=0$ are listed in Tab. \ref{tab:Outer-Peak qnm_data}, where $n=3$ mode corresponds to the WKB-identified OPS mode.
By tuning $t_0$, we see that the PS mode (red solid curve in Fig. \ref{fig:Outer-Peak-energy-fraction1})  is the most dominant component of the signal. Only at very late time, say around {$t_0 \approx 123M$}, the fundamental echo mode ($n=0$) finally gain dominance. This demonstrates a clear temporal competition between the families. Noted that this competition cannot be realized from the frequency-domain analysis in Fig. \ref{fig:QNMs-curve}, where with the same parameters, the echo family always occupies the low-lying states.

To evaluate the influence of the perturbation source on this competition, we also investigate the system's response to a wider initial Gaussian wave packet, as shown in Fig.~\ref{fig:Outer-Peak-energy-fraction2}. Comparing to Fig.~\ref{fig:Outer-Peak-energy-fraction1}, the competition still exists as expected, only the dominance of the PS family last longer, saying approximately {$134M$}. This finding is significant, as it implies that information about the perturbation source is encoded not only in the excitation  of individual QNMs but also in the hierarchical structure and competitive dynamics of the entire perturbation spectrum.

\begin{figure}[h]
    \centering
\subfigure[~Time-domain waveform]
{\includegraphics[width=1.0\linewidth]{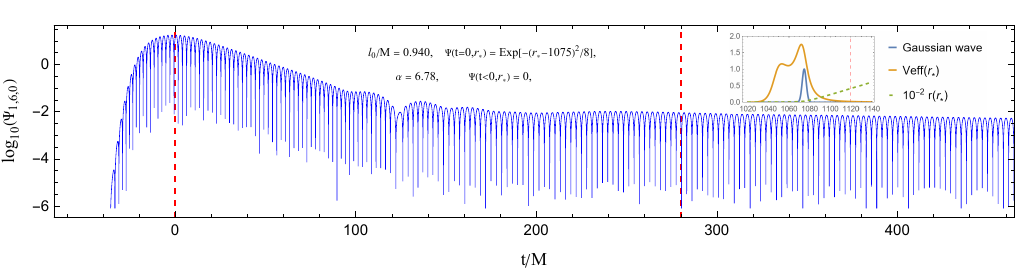}\label{fig:Outer-Peak-Dominant-waveform}}\\
\subfigure[]
{\includegraphics[width=0.4\linewidth]{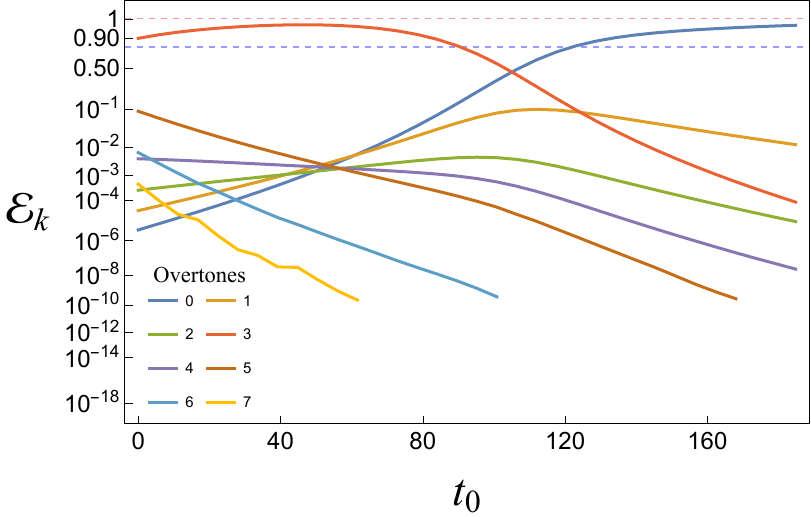}\label{fig:Outer-Peak-energy-fraction1}}\hspace{0.5cm}
\subfigure[]
{\includegraphics[width=0.4\linewidth]{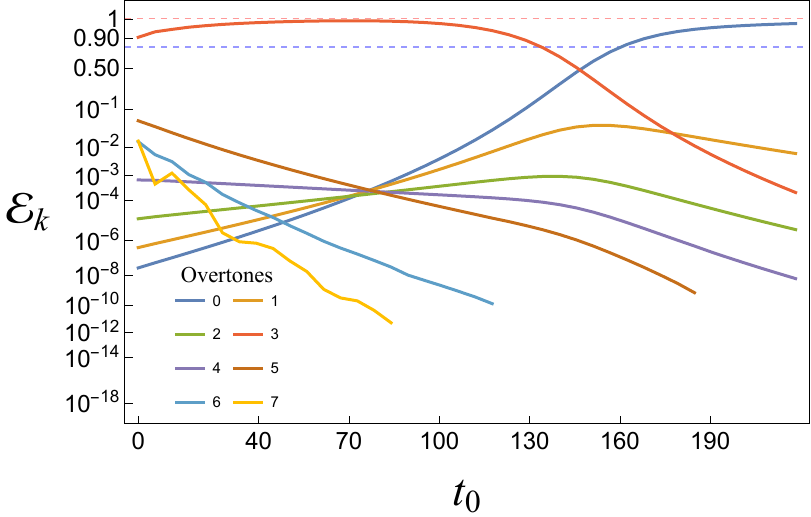}\label{fig:Outer-Peak-energy-fraction2}}~
\caption{{Time-domain analysis for the Outer-Peak dominant case:} {\bf (a)} shows the time-domain waveform, in which the inset shows the outer-peak dominant potential, the initial Gaussian wave packet, and the observer's position (vertical dashed line). The main plot's vertical red dashed line indicates the position of the observation window at $t_0=0$, with a fixed length of $T=280M$. {\bf (b)} and {\bf (c)} show the windowed energy fraction $\mathcal{E}_k$ for each mode as a function of the window's start time $t_0$. The red and blue dashed lines indicate $\mathcal{E}_k=1$ and $\mathcal{E}_k=0.8$, respectively. The latter marks the onset of the dominant mode.  Curves are truncated when the relative error {between the extracted frequency and pseudospectral method result} exceeds 0.7\%. {\bf (b)} is sourced from the initial Gaussian wave $\Psi(t=0,r_{\star})=\exp[-(r_{\star}-1075)^{2}/8]$ following {\bf (a)}, while {\bf (c)} is sourced from a wider wave packet: $\Psi(t=0,r_{\star})=\exp[-(r_{{\star}}-1075)^{2}]/72]$.}
\label{Fig:Outer-Peak Dominant case}
\end{figure}

\begin{table}[h]
  \centering
     \begin{tabular}{|c| c |c| c|}\hline
            \textbf{n} & \textbf{QNMs (MP)} & \textbf{Error} & \textbf{Complex Coefficients} \\\hline
            0 & $\pm$1.075170 - 2.817865$\times10^{-3}$ i & 1.9$\times10^{-7}$ & 0.001916 $\mp$ 0.009272 i \\\hline
            1 & $\pm$1.124298 - 2.072975$\times10^{-2}$ i & 2.0$\times10^{-7}$ & 0.020236 $\mp$ 0.055335 i \\\hline
            2 & $\pm$1.181248 - 4.559484$\times10^{-2}$ i & 2.9$\times10^{-7}$ & -0.034281 $\mp$ 0.202447 i \\\hline
            3 & $\pm$1.317639 - 6.193367$\times10^{-2}$ i & 4.2$\times10^{-7}$ & 8.701413 $\mp$ 7.564879 i \\\hline
            4 & $\pm$1.243573 - 6.943590$\times10^{-2}$ i & 2.4$\times10^{-7}$ & -0.629878 $\mp$ 0.546469 i \\\hline
            5 & $\pm$1.316021 - 9.719832$\times10^{-2}$ i & 1.4$\times10^{-5}$ & -1.164516 $\pm$ 3.578147 i \\\hline
            6 & $\pm$1.369064 - 0.140374 i & 4.4$\times10^{-4}$ & 0.182142 $\pm$ 0.910646 i \\\hline
            7 & $\pm$1.413892 - 0.175382 i & 5.8$\times10^{-3}$ & 0.086076 $\pm$ 0.164522 i \\\hline
        \end{tabular}
        \caption{QNMs and complex coefficients $\{\tilde{h}_k\}$ extracted at $t_0=0$ in Fig. \ref{fig:Outer-Peak-Dominant-waveform}. The error is the relative difference with the pseudospectral method result. The $n=3$ mode (highlighted in red in Fig. \ref{fig:Outer-Peak-energy-fraction1}) corresponds to the WKB-identified PS mode.}
        \label{tab:Outer-Peak qnm_data}
\end{table}

\item \textbf{Inner-Peak Dominant Case:}
The profile of effective potential in this scenario corresponds to the last second panel of Fig.~\ref{fig:potential-horizontal-traversal} and  the inset of Fig. \ref{fig:Inner-Peak-Dominant-waveform} in tortoise coordinates.
The typical results in this case are shown in Fig. \ref{Fig:Inner-Peak Dominant case} and Tab. \ref{tab:Inner-Peak qnm_data} for $\alpha=7.30$ and  $l_0/M=0.940$.

Fig.~\ref{fig:Inner-Peak-energy-fraction1} shows that the PS mode (purple curve, $n=4$) is excited with a very low initial energy fraction. Consequently, as the analysis window slides, this mode's contribution diminishes rapidly, leading to its anomalous disappearance from the detectable signal even before the higher-overtone echo mode ($n=5$). In contrast, when the system is perturbed by the narrower packet in Fig. \ref{fig:Inner-Peak-energy-fraction2}, the {early} energy fraction of the PS mode is significantly boosted. As a result, it persists longer in the signal and decays in the expected hierarchical order, after the $n=5$ mode.

This suggests that, in the present scenario, not only is the competition among different QNM families highly sensitive to the specific perturbation source, but the individual family also exhibit  different sensitivities to distinct perturbation source, since it alters the initial energy distributed into higher-frequency modes, resulting an observable effect at late time. Our findings suggest that future ringdown simulations could potentially leverage the differential response of competing QNM families to extract valuable information about the  perturbative sources.

\begin{figure}[ht!]
    \centering
\subfigure[]
{\includegraphics[width=1.0\linewidth]{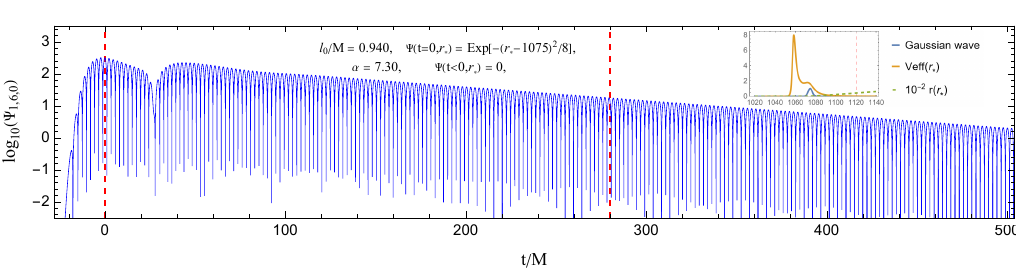}\label{fig:Inner-Peak-Dominant-waveform}}\\
\subfigure[]
{\includegraphics[width=0.4\linewidth]{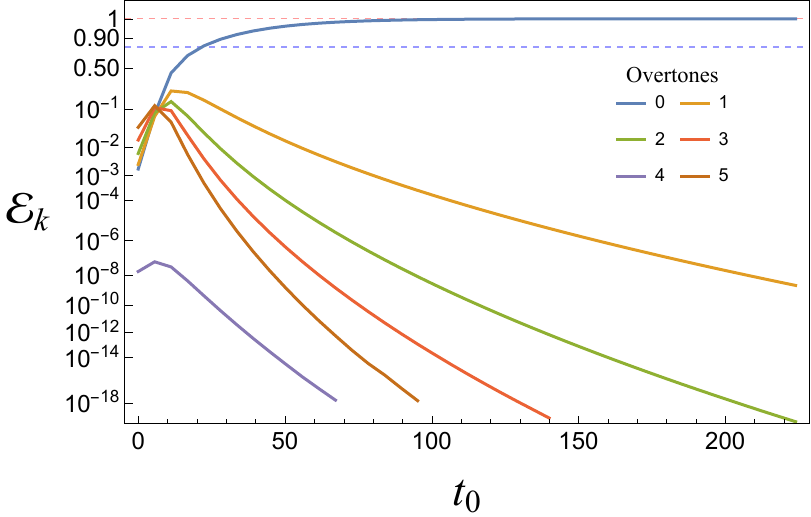}\label{fig:Inner-Peak-energy-fraction1}}\hspace{0.5cm}
\subfigure[]
{\includegraphics[width=0.4\linewidth]{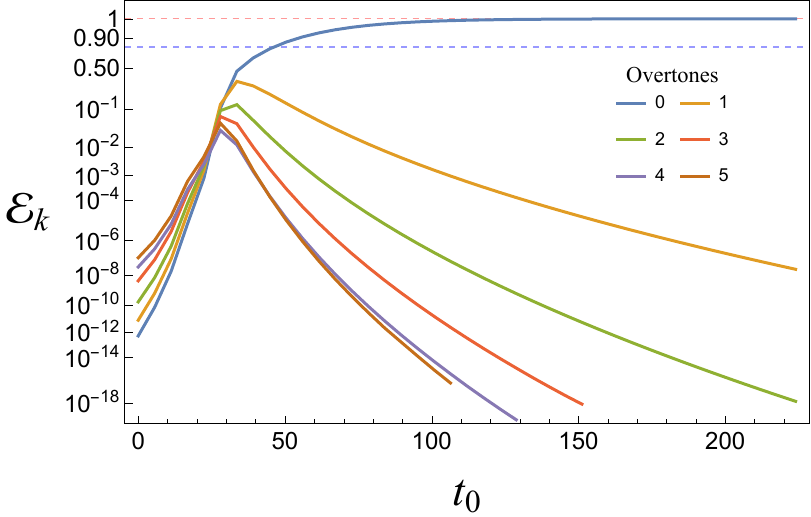}\label{fig:Inner-Peak-energy-fraction2}}~
\caption{Time-domain analysis for the Inner-Peak dominant case with $\alpha=7.30$ and $l_0/M=0.940$. Other settings are similar to those in Fig. \ref{Fig:Outer-Peak Dominant case}, but here {\bf (c)} is sourced from a narrower wave packet: $\Psi(t=0,r_{\star})=\exp[-9(r_{\star}-1075)^{2}/8]$.}
\label{Fig:Inner-Peak Dominant case}
\end{figure}

\begin{table}[ht!]
  \centering
     \begin{tabular}{|c| c |c| c|}\hline
            \textbf{n} & \textbf{QNMs (MP)} & \textbf{Error} & \textbf{Complex Coefficients} \\\hline
            0 & $\pm$1.352357 - 9.864327$\times10^{-3}$ i & 3.0$\times10^{-7}$ & 123.816399 $\mp$ 46.939802 i\\\hline
            1 & $\pm$1.437273 - 5.531763$\times10^{-2}$ i & 3.2$\times10^{-7}$ & -150.751640 $\mp$ 231.170144 i\\\hline
            2 & $\pm$1.538542 - 0.114471 i & 1.1$\times10^{-6}$ & 119.949436 $\pm$ 391.803857 i\\\hline
            3 & $\pm$1.649906 - 0.177713 i & 4.5$\times10^{-5}$ & -95.800200 $\mp$ 533.850741 i\\\hline
            4 & $\pm$2.815066 - 0.227731 i & 9.4$\times10^{-6}$ & -0.146441 $\mp$ 0.228795 i\\\hline
            5 & $\pm$1.768423 - 0.242498 i & 2.5$\times10^{-3}$ & 117.626404 $\pm$ 609.200665 i\\\hline
        \end{tabular}
        \caption{QNMs and complex coefficients $\{\tilde{h}_k\}$ extracted at $t_0=0$ in Fig. \ref{fig:Inner-Peak-Dominant-waveform}. The error is the relative difference with the pseudospectral method result. The $n=4$ mode  corresponds to the WKB-identified PS mode.}
       \label{tab:Inner-Peak qnm_data}
\end{table}

    \item \textbf{Equal-Height Peaks Case:}
In this scenario, the two potential peaks are of comparable height (see the fourth panel of Fig. \ref{fig:potential-horizontal-traversal} and the inset of Fig. \ref{fig:equal-height peaks waveform}), which represents the most trapping mechanism and the perturbation spectrum is almost dominated by the echo family. This allows us to study the dynamic competition \textit{within} the echo family itself. The results are shown in Fig. \ref{fig:equal-height peaks waveform} and Tab. \ref{tab:equal-height peaks qnm_data} with $\alpha=6.88$ and $l_0/M=0.940$. Comparing to the previous cases, the echo signal in Fig. \ref{fig:equal-height peaks waveform} is explicit.

As illustrated in Fig. \ref{fig:equal-height peaks energy fraction1}, the energy fraction reveals an obvious competition among the different overtones of echo QNM family. At very late time ({$t_0 \approx 112M$}), the $n=2$ mode (green curve) surprisingly emerges as the dominant component of the signal. Although it eventually cedes its dominance to the lower overtones, for a finite observation window, an observer would perceive this $n=2$ excited state as the dominant mode. These findings mean that in ringdown analysis, even at what might be considered as  late time, higher-overtone modes can still be dominant, {due to the fact that}  the late-time portion of the signal \textit{within the window} does not necessarily correspond to the late-time stage of the \textit{entire ringing process}.   Thus, when dealing with long-lived modes, one must be particularly cautious in defining what constitutes the late time in the rindown process, especially, in context of long-live-mode detection, one cannot simply fit the ``late-time signal within the observation window'' by sorely applying the fundamental mode.

This picture becomes even more intricate when we consider a wider initial wave packet, see Fig. \ref{fig:equal-height peaks energy fraction2}. It shows that the $n=2$ mode can still achieve dominance near a local peak where $t_0 \approx 106M$, but  then  begin yielding to the $n=1$ mode . The comparison in this case highlights the conclusion that the late-time behaviors of long-lived modes are highly sensitive to the initial properties of the perturbation source, which is in contrast to the shorter-lived PS modes generated by a single-peak potential. For PS modes, the late-time ringdown signal is generally considered to be insensitive to the initial configuration and is dominated by the fundamental mode~\cite{Berti:2005ys,LIGOScientific:2016lio,Gossan:2011ha}. So, we argue that the different responses of the echo overtones to the source could  serve as a potential new channel for extracting information about the astrophysical origin of the perturbation from future gravitational wave observations.

\begin{figure}[h]
    \centering
\subfigure[]
{\includegraphics[width=1.0\linewidth]{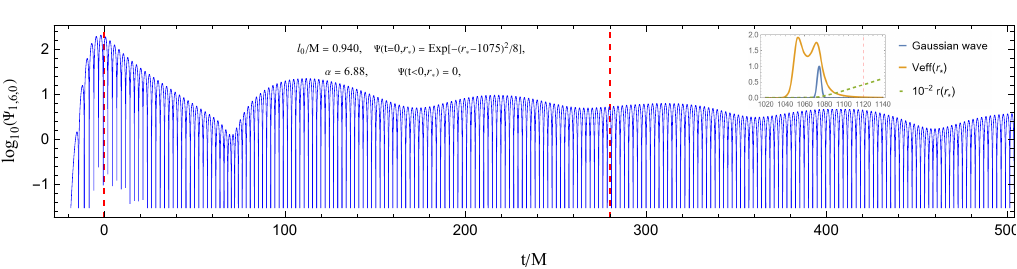}\label{fig:equal-height peaks waveform}}\\
\subfigure[]
{\includegraphics[width=0.4\linewidth]{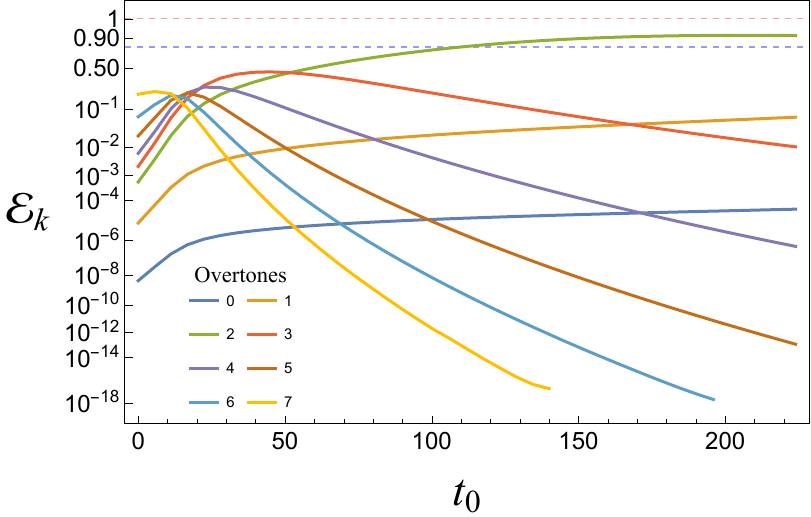}\label{fig:equal-height peaks energy fraction1}}\hspace{0.5cm}
\subfigure[]
{\includegraphics[width=0.4\linewidth]{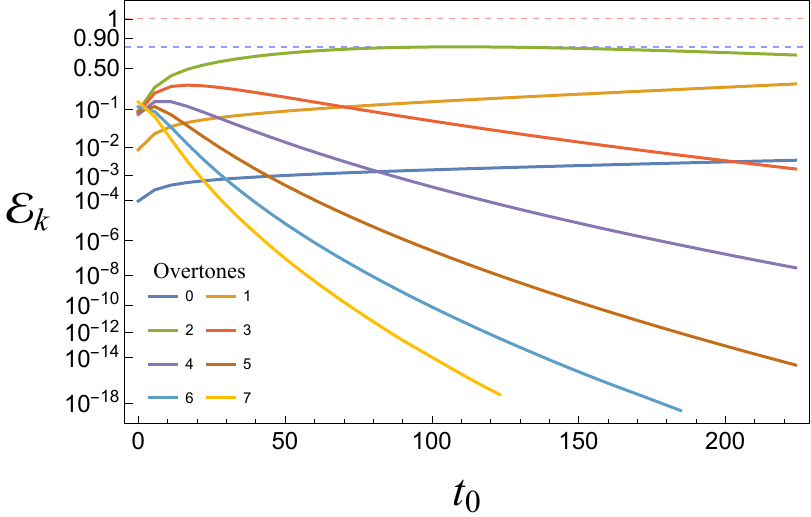}\label{fig:equal-height peaks energy fraction2}}~
\caption{Time-domain analysis for the Equal-Height Peaks case with $\alpha=6.88$ and $l_0/M=0.940$. Other settings are similar to those in Fig. \ref{fig:Outer-Peak-Dominant-waveform}.}
\end{figure}

\begin{table}[h]
  \centering
     \begin{tabular}{|c| c |c| c|}\hline
            \textbf{n} & \textbf{QNMs (MP)} & \textbf{Error} & \textbf{Complex Coefficients} \\\hline
            0 & $\pm$1.188047 - 2.628978$\times10^{-6}$ i & 2.3$\times10^{-7}$ & -0.007604 $\pm$ 0.014210 i \\\hline
            1 & $\pm$1.261862 - 2.056832$\times10^{-4}$ i & 2.5$\times10^{-7}$ & -0.285697 $\pm$ 0.625067 i \\\hline
            2 & $\pm$1.324950 - 3.774128$\times10^{-3}$ i & 2.7$\times10^{-7}$ & -5.510310 $\pm$ 5.888495 i \\\hline
            3 & $\pm$1.379934 - 1.692077$\times10^{-2}$ i & 2.8$\times10^{-7}$ & -20.906758 $\pm$ 18.881230 i \\\hline
            4 & $\pm$1.430182 - 4.144860$\times10^{-2}$ i & 2.9$\times10^{-7}$ & 16.786243 $\pm$ 59.969066 i \\\hline
            5 & $\pm$1.483314 - 7.812370$\times10^{-2}$ i & 4.1$\times10^{-7}$ & 58.581118 $\pm$ 109.418735 i \\\hline
            6 & $\pm$1.541224 - 0.120325 i & 6.2$\times10^{-6}$ & 206.524788 $\pm$ 33.493797 i \\\hline
            7 & $\pm$1.602213 - 0.166283 i & 1.6$\times10^{-4}$ & 202.887553 $\mp$ 243.668596 i \\\hline
        \end{tabular}
        \caption{QNMs and complex coefficients $\{\tilde{h}_k\}$ extracted at $t_0=0$ in Fig. \ref{fig:equal-height peaks waveform}. The error is the relative difference with the pseudospectral result. All the modes are contributed solely from the echo family.}
        \label{tab:equal-height peaks qnm_data}
\end{table}

\end{enumerate}

%======================================%
%<<<<<<<<<<< Conclusions >>>>>>>>>>>>>%
%======================================%
\section{Conclusions and discussion}\label{sec: conclusions}
In this work, we performed a comprehensive investigation into the eletromagnetic perturbation dynamics of hairy Schwarzschild black holes, which admit both single and double photon spheres. Instead of merely identifying phenomena such as echoes, we establish a physical classification of the QNM spectrum and reveal a new layer of dynamics in the time-domain ringdown signal. Our analysis indicates a rich global structure of QNM spectra arising from the complex structure of the effective potential for this black hole. Our main findings are summarized as follows.

\begin{itemize}
    \item
 With the use of suite of parallel methods, we identified that the QNM spectrum in this scenario generates two physical families: a \textit{photon sphere family} and an \textit{echo family}. We proposed that the WKB approximation can serve as an effective marker for these families. Moreover, we argue that the \textbf{degree of localization of the QNM eigenfunction} in the vicinity of the potential barriers can be an unambiguous criterion for their classification, from which we can physically distinguish among the highly delocalized \textit{scattering resonances} (the PS family), highly localized \textit{quasi-bound states} (the Echo family), and  intermediate \textit{diffraction-trapped states} observed near border of sing peak and double peaks. In particular, it is seen that from the frequency domain, the strong trapping regime, where the two potential peaks are of comparable height, the traditional PS family predicted by geometric optics is effectively suppressed or absent from the spectra.

    \item
    Motivated by the presence of extremely long-lived modes and in this case an observational data may be inherently windowed, we developed a \textit{windowed energy analysis framework} to quantitatively study the ringdown signal, which reveals a dynamic competition for dominance among different QNM families. Our results show that the dominant mode within a given analysis window, even one starting at what is considered a ``late" time, is not necessarily the fundamental mode of the entire spectra. It could instead be a mode from the PS family or even a high-overtone mode from the echo family, depending on the window's start time.

    Moreover, our studies also indicate  that the competitive dynamic for dominance is profoundly sensitive to the properties of the initial perturbation source. In particular, the initial energy distribution imparted by the source dictates the hierarchy and evolution of the competing modes. This feature is different from the short-live signal in GR, in which the black hole's  late-time ringdown is independent of the initial excitation \cite{LIGOScientific:2016lio,Gossan:2011ha}.
\end{itemize}

Our current studies can be extended into the gravitational perturbations to analyze the gravitational ringdown signal. We shall argue that our preliminary results can provide significant insights on the observations and analysis of ringdown signal. Firstly, our results indicate that in some cases, higher overtone may become dominant at late time within a certain observational windows. It further confirms that traditional consideration that the observed signal decays into the fundamental mode, might be complemented by an alternative strategy \cite{Giesler:2019uxc,Oshita:2021iyn}. Especially, for the system producing long-live signals, developing tailored fitting models, such as those based on a single, pure excited state, is motivated for a more faithful representation of the waveform in the window. Secondly,  we suggest that the time-dependent energy competition among QNM families may itself serve as a distinctive observable. In particular, the temporal re-ordering of dominant modes during ringdown window could provide potentially robust way of differing hairy black holes from their counterpart in GR, helping to further test no-hair theorem with ringdown overtone signal \cite{Bhagwat:2019dtm}. The last but not least, the spectral evolution of the ringdown, including the dynamic {distribution} and underlying transfer of energy among QNM families,  encodes rich information of both the spacetime geometry (the profile of perturbation potential) and mechanism driving the perturbation. This raises the possibility of using the ringdown's temporal structure to probe more information about the wave sources.

Additionally, it is known  that in the eikonal limit of GR and many other theories,  $\text{Re}(\omega)$ corresponds to the orbital frequency of an unstable null geodesic at the PS, while $\text{Im}(\omega)$ corresponds to its Lyapunov exponent, which quantifies the characteristic timescale of the orbit's instability \cite{Cardoso:2008bp,Dolan:2010wr}.
However, here in the case of a double-peak potential, particularly when the peaks are of similar height, the QNM spectrum is almost governed by the echo family. This corresponds to the eigenfunction configurations in the middle panel of Fig. \ref{fig:eigenfunctions}. The energy levels of echo modes $\text{Re}(\omega_{n})^{2}$ are high enough to reach the potential's peak, which possesses an energy of an unstable orbit on the PS in the geometric-optics picture. Nevertheless, their decay rates given by $\text{Im}(\omega_n)$ are determined by their overtone numbers $n$ within the echo family, which is different from the decay rate predicted by the WKB approximation at the local maximum of the potential, related to the Lyapunov exponent. In a word, since for black holes with double PS, the QNMs corresponding to the PS are either absent or extremely suppressed, therefore, the link inferring the geometric-optics properties like the Lyapunov exponent from the QNM spectra breakdowns.
It suggests that instead of the geometric-optics approximation, the QNM structure of these black holes could require a more rigorous framework, which can fully incorporates the wave effects of trapping and resonant tunneling. This deserves further efforts. It is worthwhile to note that
the authors of \cite{Duran-Cabaces:2025sly} recently attempted to infer the Lyapunov exponent from the properties of extra rings formed by double-peak potentials, but found significant discrepancies with QNM calculations. Our current study could offer a parallel physical explanation for this discrepancy.

%======================================%
%<<<<<<<<<< Acknowledgement >>>>>>>>>>>%
%======================================%
\section*{Acknowledgement}
This work was supported in part by the National Natural Science Foundation of China with Grants Nos. 12375054 and 12505067, and the Postgraduate Research \& Practice Innovation Program of Jiangsu Province with Grant No. KYCX25\_3923.
	
{{\bf Data \& code Availability:} Data and code to obtain the data in this work are made publicly available on \href{https://github.com/Yang070201/arxiv251002033}{GitHub}}.

\appendix

\section{Hairy Schwarzschild black hole from gravitational decoupling}
\label{appendix:background}

In this appendix, we shall briefly review how the hairy Schwarzschild black hole can be constructed by the GD algorithm, which was developed for analytically finding a deformed solution to Einstein's field equation with additional source term~\cite{Ovalle:2018gic,Ovalle:2017wqi,Ovalle:2020kpd}. The methodology can be started from taking the additional source as a decoupled subpart of a total energy-momentum tensor $ \tilde{T}_{\mu\nu}$, and the Einstein equations take the form
\begin{eqnarray}\label{eq:decoupled T}
    \tilde{G}_{\mu\nu}\equiv R_{\mu\nu}-\frac{1}{2}Rg_{\mu\nu}=\kappa\,\tilde{T}_{\mu\nu} \equiv \kappa(T_{\mu\nu}+\Theta_{\mu\nu})\, ,
\end{eqnarray}
where $T_{\mu\nu}$ represents the energy-momentum tensor corresponding to a known seed GR solution, while $\Theta_{\mu\nu}$ describes the contributions from new matter fields or additional gravitational sectors. The simplest seed solution is the Schwarzschild vacuum, given by  $T_{\mu\nu}=0$, then one can consider a deformed Schwarzschild solution to \eqref{eq:decoupled T} with the ansatz
\begin{eqnarray}\label{eq:hairy ansatz}
    \mathrm{d}s^2=-\Big( 1-\frac{2\mathsf{M_s}}{r}\Big)\mathrm{e}^{\alpha\, k(r)}\mathrm{d}t^2+\left( 1-\frac{2\mathsf{M_s}}{r}\right)^{-1}\mathrm{e}^{-\alpha\, k(r)}\mathrm{d}r^2+r^2\mathrm{d}\Omega^2\, ,
\end{eqnarray}
where $\mathsf{M_s}$ represents the Schwarzschild mass and $k(r)$ is a function to be determined. The parameter $\alpha$ quantifies the deformation strength. $\alpha=0$ indicates no additional source $\Theta_{\mu\nu}$, reverting the solution to the original seed metric without deformation. It is worth noting that this ansatz becomes trivial deformation when the function $k(r)$ is satisfied with
\begin{eqnarray}
    \quad \mathrm{e}^{\alpha\, k(r)}=1-\frac{\alpha\, c_0}{r}\left(1-\frac{2\mathsf{M_s}}{r}\right)^{-1}~~\text{with}~~c_0=\text{Const.}\, ,
\end{eqnarray}
because it leads to
\begin{eqnarray}
    -g_{tt}=g_{rr}^{-1}=1-\frac{2\mathsf{M_{s}}+\alpha\, c_0}{r}=1-\frac{2\mathsf{M'_s}}{r}\, ,
\end{eqnarray}
which leaves the original Schwarzschild geometry unaffected, and only the mass parameter will be changed.

The key point of the GD method lies in its approach to impose the effect of the deformation source $\Theta_{\mu\nu}$ via a specific geometric deformation. This operation cleverly splits the originally highly-coupled Einstein's field equations into two separate parts. For the seed Schwarzschild solution, one part consists of the original Einstein's equations for the seed spacetime sourced by $T_{\mu\nu}$, i.e., $G^{\mu}{}_{\nu}(r,\mathsf{M_s})=T^{\mu}{}_{\nu}=0$. The other is a set of ``quasi-Einstein'' equations that describe the additioanl source  and its influence on the background geometry, i.e., $\alpha~\mathcal{G}^{\mu}{}_{\nu}(r,\mathsf{M_s},\alpha~k(r))=\Theta^{\mu}{}_{\nu}$. The Einstein tensor $\tilde{G}^{\mu}{}_{\nu}$ is consist of $G^{\mu}{}_{\nu}(r,\mathsf{M_s})$ and $\alpha~\mathcal{G}^{\mu}{}_{\nu}(r,\mathsf{M_s},k(r))$. In order to find a specific, physically intuitive hairy black hole solution in this framework, one must solve the ``quasi-Einstein'' equation to determine $k(r)$. To this end, additional physical constraints must be imposed on the additional source. The authors of~\cite{Ovalle:2020kpd} imposed the following two key conditions to constrain the additional source. Firstly, the deformed solution should have a well-defined horizon with radius $r_h$ such that $g_{tt}(r_{h})=g_{rr}^{-1}(r_{h})=0$, which leads to an equation of state from Einstein's equation
\begin{eqnarray}\label{eq:equation of state}
    \tilde{p}_{r}=-\tilde{\rho}\quad \text{or}\quad \Theta^{r}{}_{r}=\Theta^{t}{}_{t}\, ,
\end{eqnarray}
where the effective density, radial pressure and tangential pressure can be taken from
\begin{eqnarray}
    \tilde{T}^{\mu}{}_{\nu}\equiv\text{Diag}\{-\tilde{\rho},\tilde{p}_{r},\tilde{p}_{t},\tilde{p}_{t}\}=T^{\mu}{}_{\nu}+\Theta^{\mu}{}_{\nu}\, .
\end{eqnarray}
Noted that in the current framework we already have $T^{\mu}{}_{\nu}=0$. Secondly, the strong energy condition is imposed to avoid exotic matter sources, which guarantees the convergence of time-like geodesics affected by the tidal force of matter sources, requiring
\begin{eqnarray}\label{eq:SEC}
    \tilde{\rho}+\tilde{p}_r+2\tilde{p}_t \ge 0\, ,~~
    \tilde{\rho}+\tilde{p}_r \ge 0\, ,~~\tilde{\rho}+\tilde{p}_t \ge 0\, .
\end{eqnarray}
Subsequently, the ``quasi-Einstein'' equations combining with the above conditions yield two second-order linear differential inequalities
\begin{eqnarray}\label{eq:GD conditions 1}
    \frac{1}{2\kappa\,r}\Big[ h''(r-2\mathsf{M_s})+2h' \Big]&=& \Theta^{\theta}{}_{\theta}\ge0 \,,\\
    \frac{1}{2 \kappa\,r^2} \Big[  h''(r-2\mathsf{M_s})r + 4h'\mathsf{M_s}-2h+2 \Big]&=& \Theta^{\theta}{}_{\theta}-\Theta^{t}{}_{t}\ge0\,,
\end{eqnarray}
where $h\equiv \mathrm{e}^{\alpha\, k(r)}$ and the prime denotes the derivative with respect to $r$. One may consider the deformation should be trivial at $r=r_{h}$ and $r\gg \mathsf{M_s}$. Since the deformed black hole admits a proper horizon $r_{h}\sim 2\mathsf{M_s}$ and a asymptotically Schwarzschild geometric, thus a simple example  emerges as
\begin{eqnarray}\label{eq:hairy sol 0}
\Theta^{\theta}{}_{\theta}=\frac{\alpha}{2\kappa\,\mathsf{M_s}^2 r}\mathrm{e}^{-r/\mathsf{M_s}}\left(r-2\mathsf{M_s}\right)\, .
\end{eqnarray}
Therefore, the deformation function sourced by \eqref{eq:hairy sol 0} can be solved from the second-order differential equation \eqref{eq:GD conditions 1}
\begin{eqnarray}\label{eq:hairy sol 1}
    h(r)=c_{1}-\alpha \frac{c_{2}-r\,\mathrm{e}^{-r/\mathsf{M_s}}}{r-2\mathsf{M_s}}=1-\left(\frac{l_0}{r}-\alpha\,\mathrm{e}^{-\frac{r}{\mathsf{M_s}}}\right)\left(1-\frac{2\mathsf{M_s}}{r}\right)^{-1},
\end{eqnarray}
where the integration constant $c_{1}=1$ has been taken for keeping asymptotical flatness and $\alpha\, c_{2}\equiv l_0$ has been defined for representing a hairy charge associated with primary hair.

Subsequently, a hairy black hole solution deformed from the Schwarzschild seed metric reads \cite{Ovalle:2020kpd}
\begin{eqnarray}
    &&\mathrm{d}s^2=-A(r)\mathrm{d}t^2+\frac{\mathrm{d}r^2}{B(r)}+C(r)(\mathrm{d}\theta^2+\sin^2\theta\mathrm{d}\phi^2)\, ,\nonumber\\
    &&\text{with}\quad   A(r)=B(r)=1-\frac{2M}{r}+\alpha\, \mathrm{e}^{-\frac{r}{M-l_{0}/2}}\equiv f(r) \quad \text{and}\quad   C(r)=r^2\, ,\label{eq:hairy sol 2a}
\end{eqnarray}
where $M=\mathsf{M_s}+l_{0}/2$ represents the asymptotic mass of the  hairy black hole in this framework. It is noted that, in general,  different choice of $\Theta_{\mu\nu}$ could result to different hairy black hole holes, but the hairy Schwarzschild black hole \eqref{eq:hairy sol 2} was obtained by considering $\Theta_{\mu\nu}$ as the stress energy tensor of an anisotropic fluid. Nevertheless, this hairy black hole still has its generality because it does not rely on specific matter fields and accommodates a wide variety of hair, such as scalar hair, tensor hair, and even fluid-like dark matter, etc.. Thus, some theoretical aspects and observational phenomena of this hairy black hole and its rotating counterpart have been widely studied. For example, the thermodynamical properties were studied in \cite{Mahapatra:2022xea}. The observational signatures through strong gravitational lensing, black hole shadow and images were explored in \cite{Afrin:2021imp,Islam:2021dyk,Ghosh:2022kit,Afrin:2021wlj,Meng:2024puu,Meng:2025ivb} which are based on the deflection of photons. Starting from the geodesic motion of timelike particle, the effects such as  precession and Lense-Thirring effect \cite{Wu:2023wld}, gravitational waves from extreme mass ratio inspirals \cite{Zi:2023omh} and quasi-periodic oscillations \cite{Liu:2023ggz} have also been examized.

\section{Reducing procedure of the master scalar with other spins}\label{appendix:master equation}
In this Appendix, we will briefly shown the reducing procedure of the master scalar \eqref{eq: master eq} for the scalar field ($s=0$) and gravitational perturbation ($s=2$).

\begin{itemize}
    \item For $s=0$, the perturbation could be sourced by an external scalar field and then be governed by the Klein-Gorden equation in the black hole background
\begin{eqnarray}\label{eq:KG equation}
	\square\Phi=\frac{1}{\sqrt{-g}}(g^{\mu\nu}\sqrt{-g}\Phi_{,\mu})_{,\nu}=0\, ,
\end{eqnarray}
    in which $g$ denotes the determinant of the background. Considering a standard separation of variables of the form
\begin{eqnarray}\label{eq: separation_of_variables}
    \Phi(t,r,\theta,\phi)=\sum_{\ell,m}\frac{\Psi_{0,\ell,m}(t,r)}{r}Y_{\ell,m}(\theta,\phi)\, ,
\end{eqnarray}
it is straightforward to obtain that the master scalar $\Psi_{0,\ell,m}$ will involve via  the master equation \eqref{eq: master eq} with $s=0$, and the standard scalar spherical harmonics $Y_{\ell,m}$ is governed by
\begin{eqnarray}
    \left[ \frac{1}{\sin\theta}\frac{\partial}{\partial \theta}\left(\sin\theta \frac{\partial}{\partial \theta}\right) + \frac{1}{\sin^2\theta}\frac{\partial^2}{\partial \phi^2} \right] Y_{\ell,m}(\theta, \phi) = -\ell(\ell+1) Y_{\ell,m}(\theta, \phi)\, .
\end{eqnarray}

   \item For $s=2$, we consider a simple case that perturbations are excited by a particle falling radially into the black hole. Due to the spherical symmetry, the resulting scalar perturbations are of pure even-parity. Consequently, the odd-parity gravitational perturbations naturally decouple and evolve independently at the linear level. Their dynamics are thus governed by a homogeneous Einstein equation, whose solutions reveal the intrinsic dynamic of the spacetime itself
\begin{eqnarray}\label{eq:linear einstein}
    G^{(1)}_{\mu\nu}[h^{\text{odd}}]&=& \left(\delta_{\mu}{}^{\alpha}\delta_{\nu}{}^{\beta}-\frac{1}{2}g_{\mu\nu}g^{\alpha\beta}\right)R^{(1)}_{\alpha\beta}[h^{\text{odd}}]=0\, ,\\
    \text{with}\quad h^{\text{odd}}_{\mu\nu}&= &
    \begin{pmatrix}
    0&0&-h_0(t,r)\,\csc\theta\,\partial_{\phi}&h_0(t,r)\,\sin\theta\,\partial_{\phi}\\
    0&0&-h_1(t,r)\,\csc\theta\,\partial_{\phi}&h_1(t,r)\,\sin\theta\,\partial_{\phi}\\
    - h_0(t,r)\,\csc\theta\,\partial_{\phi}&-h_1(t,r)\,\csc\theta\,\partial_{\phi}&0&0\\
    h_0(t,r)\,\sin\theta\,\partial_{\phi}&h_1(t,r)\,\sin\theta\,\partial_{\phi}&0&0
 \end{pmatrix}Y_{\ell, m}\, ,
\end{eqnarray}
where $G^{(1)}_{\mu\nu}$ and $R^{(1)}_{\alpha\beta}$ denote the linear Einstein tensor and linear Ricci tensor. The odd-parity metric perturbation, $h^{\text{odd}}_{\mu\nu}$, is constructed from the Regge-Wheeler basis functions and can be expressed in terms of two radial functions, $h_0(t,r)$ and $h_1(t,r)$. Following the standard procedure, we introduce a master scalar function $\Psi^{\text{odd}}_{2,\ell,m}$ to encapsulate their dynamics. A convenient definition relates the master scalar to the metric functions as follows:
\begin{eqnarray}
    h_{1}&=&\sqrt{\frac{C}{A\,B}}\,\Psi^{\text{odd}}_{2,\ell,m}=\frac{r}{f} \Psi^{\text{odd}}_{2,\ell,m}\, ,\\
    \dot{h}_{0}&=&\sqrt{A\,B}\,(\sqrt{C}\,\Psi^{\text{odd}}_{2,\ell,m})'=f\,(r\,\Psi^{\text{odd}}_{2,\ell,m})'\, ,
\end{eqnarray}
the dynamical components $G^{(1)}_{31}[\Psi^{\text{odd}}_{2,\ell,m}]=G^{(1)}_{13}[\Psi^{\text{odd}}_{2,\ell,m}]$ of linear Einstein's equation \eqref{eq:linear einstein} can be finally reduced to the master equation \eqref{eq: master eq} with $s=2$.

\end{itemize}

\section{Hyperboloidal framework of hairy Schwarzschild black hole}\label{appendix:hyperboloidal_framework}
In this appendix, we employ the so-called ``scri-fixing technique'' to construct the hyperboloidal coordinate~\cite{Zenginoglu:2007jw,PanossoMacedo:2023qzp}. Specifically, the hyperboloidal coordinate transformations are given by
\begin{eqnarray}\label{hyperboloidal_transformation}
    t=r_{h}(\tau-H(\sigma))\, ,\quad r=\frac{r_{h}}{\sigma}\, ,
\end{eqnarray}
where the length scale has been chosen by the event horizon $r_{h}$. There only exists one horizon for the double-peak effective potential, and the horizon will be $\sigma=1$. From \eqref{eq:hairy sol 2a}, in terms of $\sigma$, the metric function can be expressed as
\begin{eqnarray}
    \mathcal{F}(\sigma)=f(r(\sigma))=1-\frac{2M}{r_{h}}\sigma+\alpha\exp\Big(-\frac{r_{h}}{(M-l_0/2)\sigma}\Big)\, .
\end{eqnarray}
The derivative of the dimensionless tortoise coordinate $x(\sigma)$ is expanded as
\begin{eqnarray}
    x^{\prime}(\sigma)=-\frac{1}{\sigma^2\mathcal{F}(\sigma)}=-\frac{1}{\sigma^2}-\frac{2M}{r_{h}}\frac{1}{\sigma}+\mathcal{O}(1)\, .
\end{eqnarray}
The leading terms contribute with the singular quantities, i.e.,
\begin{eqnarray}
    x_0(\sigma)=\frac{1}{\sigma}-\frac{2M}{r_{h}}\ln\sigma\, .
\end{eqnarray}
At the event horizon $\sigma=1$, the integration around the horizon $\sigma=1$ yields
\begin{eqnarray}
    x_{h}(\sigma)=\frac{1}{K_{h}(1)}\ln|\sigma-1|\, ,\quad \text{with}\quad
    K_{h}(1)=\frac{2M}{r_{h}}+\frac{2\alpha r_{h}}{l_0-2M}\exp\Big(\frac{2r_{h}}{l_0-2M}\Big)\, .
\end{eqnarray}
The regular piece $x_{\text{reg}}(\sigma)$ is satisfied with
\begin{eqnarray}
    x^{\prime}_{\text{reg}}(\sigma)=x^{\prime}(\sigma)-x^{\prime}_{h}(\sigma)-x^{\prime}_0(\sigma)\, .
\end{eqnarray}

Therefore, derivatives of two height functions $H^{\text{in-out}}(\sigma)$ and $H^{\text{out-in}}(\sigma)$ associated with $\sigma$ are
\begin{eqnarray}\label{height_function_in_out}
    \frac{\mathrm{d}H^{\text{in-out}}(\sigma)}{\mathrm{d}\sigma}=-\frac{1}{\sigma^2\mathcal{F}(\sigma)}+\frac{2}{\sigma^2}+\frac{4M}{r_{h}}\frac{1}{\sigma}\, ,
\end{eqnarray}
and
\begin{eqnarray}\label{height_function_out_in}
    \frac{\mathrm{d}H^{\text{out-in}}(\sigma)}{\mathrm{d}\sigma}=\frac{1}{K_{h}(1)}\frac{2}{\sigma-1}+\frac{1}{\sigma^2\mathcal{F}(\sigma)}\, .
\end{eqnarray}
Note that when $\alpha=0$, i.e., the case of the Schwarzschild black hole, two height functions are equal. Ref. \cite{PanossoMacedo:2023qzp} tells us how to use the approach presented in it to obtain the hyperboloidal coordinate in a single chart. For some black holes, such as Schwarzschild black hole, RN black hole~\cite{Destounis:2021lum}, quantum corrected Schwarzschild black hole~\cite{Cao:2024oud}, Hayward black hole~\cite{Wu:2024ldo}, and Boulware-Deser-Wheeler black hole~\cite{Cao:2024sot}, a single hyperboloidal chart can indeed be achieved. But things turned out contrary to one's wishes. We find that within some certain parameters, applying the method provided in~\cite{PanossoMacedo:2023qzp} to this hairy black hole will result in the failure of the construction of hyperboloidal framework using the height function Eq. (\ref{height_function_in_out}) and Eq. (\ref{height_function_out_in}). Similar case will happen in the black hole with halo~\cite{PanossoMacedo:2023qzp}. Hence, we need to give some complements for the hyperboloidal framework.

\begin{figure}[htbp]
	\centering
	\includegraphics[width=0.45\textwidth]{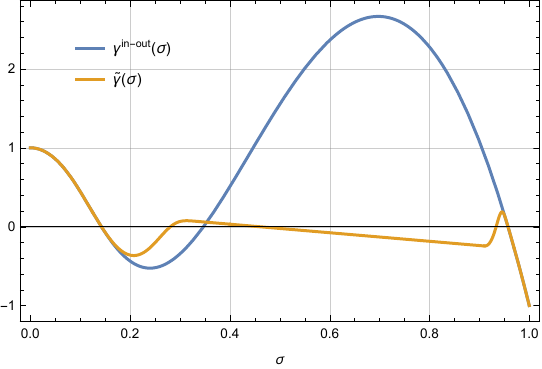}\hspace{0.5cm}
    \includegraphics[width=0.5\textwidth]{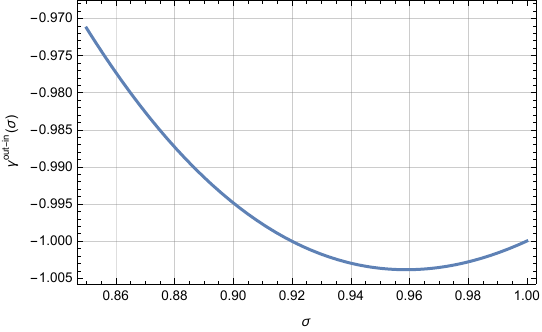}
	\caption{The functions $\gamma^{\text{in-out}}(\sigma)$, $\tilde{\gamma}(\sigma)$ and $\gamma^{\text{out-in}}(\sigma)$ are shown. For the left panel with the in-out strategy, the parameters are chosen as $M=1$, $l_0=1.02M$ and $\alpha=9.312$. For the right panel with the out-in strategy, the parameters are chosen as $M=1$, $l_0=1.02M$ and $\alpha=7.312$. The above parameter selections are from~\cite{Yang:2024rms}.}
	\label{gamma_counterexample}
\end{figure}

It arises from demanding $g_{\sigma\sigma}>0$, so that the correct signature of metric can be still maintained. The condition $g_{\sigma\sigma}>0$ yields that $|\gamma|<1$, which means that the hypersurfaced $\tau=\text{constant}$ keep spacelike in the concerned domain $\sigma\in(0,1)$ from a geometric perspective. The function $\gamma(\sigma)$ is given by
\begin{eqnarray}
    \gamma(\sigma)=\frac{\mathrm{d}H(\sigma)}{\mathrm{d}\sigma}p(\sigma)\, .
\end{eqnarray}
However, after our test, it is found that both strategies derived from~\cite{PanossoMacedo:2023qzp} will encounter situations where condition $|\gamma|<1$ is not met. As some typical examples, in the Fig. \ref{gamma_counterexample}, we choose some parameters to show that $|\gamma|>1$ appears in the part of the interval $(0,1)$. Although Ref. \cite{PanossoMacedo:2023qzp} suggests that the out-in strategy is better, what we want to say here is that the out-in strategy is even worse within such chosen of parameters. The situation in the left panel of the Fig. \ref{gamma_counterexample} can still be remedied, while the situation in the right panel of the Fig. \ref{gamma_counterexample} is already regarded as hopeless. One key difference between these two is the sign of the derivative of function $\gamma(\sigma)$ at the endpoint of the interval including $\sigma=1$. We have $\mathrm{d}\gamma^{\text{in-out}}(\sigma)/\mathrm{d}\sigma(\sigma=1)<0$ but $\mathrm{d}\gamma^{\text{out-in}}(\sigma)/\mathrm{d}\sigma(\sigma=1)>0$ in these examples. Therefore, one can find some interval $(\sigma_0,1)$, in which the function $\gamma^{\text{in-out}}(\sigma)$ is satisfied with $|\gamma|<1$ in such interval. However, it is not achievable for the function $\gamma^{\text{out-in}}(\sigma)$ at this time.

Now, we provide remedies if a situation similar to the left panel of Fig. \ref{gamma_counterexample} occurs. Motivated by~\cite{Zenginoglu:2007jw}, hyperboloidal foliations can be constructed that are spacelike in the interior without affecting the QNM spectra. To accomplish this objective, it is easy to construct the new function $\tilde{\gamma}(\sigma)$ (yellow line in left panel of Fig. \ref{gamma_counterexample}), which is satisfied with $|\tilde{\gamma}(\sigma)|<1$, where the asymptotic behaviors of $\gamma^{\text{in-out}}(\sigma)$ and $\tilde{\gamma}(\sigma)$ are the same. In practice, if we encounter the above problems, we will adopt $\tilde{\gamma}(\sigma)$ instead of $\gamma(\sigma)$.

Using the obtained hyperboloidal coordinates (\ref{hyperboloidal_transformation}), we recast Eq. (\ref{eq: master eq}) into a form of a hyperbolic equation
\begin{eqnarray}\label{PDE}
   \partial_\tau u=\mathrm{i}Lu\, ,\quad
   L=\frac{1}{\mathrm{i}}
   \begin{pmatrix}
   0    &   1   \\
   L_1  &   L_2
   \end{pmatrix}\, ,\quad
   u=\begin{pmatrix}
   \Psi \\
   \Pi
   \end{pmatrix}\, ,
\end{eqnarray}
where $\Pi=\partial_\tau \Psi$ is introduced to reduce the time derivative order in the equation, and the explicit expressions of differential operators $L_1$ and $L_2$ are
\begin{eqnarray}
    L_{1}=\frac{1}{w(\sigma)}\Big[\partial_{\sigma}(p(\sigma)\partial_{\sigma})-q_s(\sigma)\Big]\, ,~~~~~~
    L_{2}=\frac{1}{w(\sigma)}\Big[2\gamma(\sigma)\partial_{\sigma}+\partial_{\sigma}\gamma(\sigma)\Big]\, ,
\end{eqnarray}
where the explicit forms of the above function are given as follows
\begin{eqnarray}
    p(\sigma)=\sigma^2\mathcal{F}(\sigma)\, ,\quad w(\sigma)=\frac{1-\gamma^2(\sigma)}{p(\sigma)}\, ,\quad q_s(\sigma)=\frac{r_h^2}{p(\sigma)}V_s(\sigma)\, .
\end{eqnarray}
Having the above first-order-in-time PDE system (\ref{PDE}), and performing the Fourier transformation in $\tau$ with $u(\tau,\sigma)\sim u(\sigma)\mathrm{e}^{\mathrm{i}\omega\tau}$, we arrive at the eigenvalue problem $Lu=\mathrm{i}\omega u$. To numerically compute the QNM spectra, we discretize the differential operator $L$ using a Chebyshev-Lobatto collocation scheme (pseudospectral method), which yields its matrix representation $\mathbf{L}$. This discretization effectively reduces the original infinite-dimensional spectral problem to a finite-dimensional algebraic eigenvalue problem
\begin{eqnarray}\label{eigenvalue_porblem}
    \mathbf{L}\mathbf{u}=\mathrm{i}\omega \mathbf{u}\, .
\end{eqnarray}
Note that we have $\omega_\tau=r_h\omega_t$, in which $\omega_\tau$ is derived in the $\tau$-coordinate and $\omega_t$ is derived in the $t$-coordinate [cf. Eq. (\ref{hyperboloidal_transformation})]. Note that when the master equation has a correction term of the velocity, the corresponding hyperbolidal framework can still work~\cite{Cao:2025qws}. Furthermore, the hyperboloidal framework for the Kerr spacetime is constructed in~\cite{PanossoMacedo:2019npm}, which is used to obtain the QNM spectra~\cite{Cai:2025irl,Assaad:2025nbv}. See also some comments on the hyperboloidal framework for black hole QNMs~\cite{Qian:2025jht}.

\section{Direct integration method}
\label{sec:appendix_direct integration}
In this appendix, we detail the semi-analytic method used to compute the QNM spectra in the frequency domain. The approach is based on finding series solutions to the master equation in the asymptotic regions near the event horizon and spatial infinity. Then matching these solutions in an intermediate region. The fundamental mathematical idea of this direct integration method was originally established by Chandrasekhar \cite{Chandrasekhar:1975}. Subsequently, this technique has been widely applied in various contexts, such as in the study of dynamical Chern-Simons gravity \cite{Wagle:2021tam, Molina:2010fb}, and the black hole with a thin shell~\cite{Laeuger:2025zgb}.

We begin with Eq. \eqref{master_equation_frequency_domian}. In the spirit of the Frobenius method, we seek solutions in the asymptotic regions by factoring out the singular behavior, which is determined by the QNM boundary conditions (purely ingoing at the horizon and purely outgoing at infinity). At the event horizon $(r \to r_h)$, the solution takes the form:
\begin{eqnarray}
    \Psi_{\text{h}}(r) = (r-r_h)^{-\frac{\mathrm{i}\,\omega}{2\kappa}}\sum^{\mathcal{H}}_{j=0}\mathfrak{h}_{j}(r-r_h)^{j}\, ,
\end{eqnarray}
and at spatial infinity $(r \to \infty)$, the solution is written as:
\begin{equation}
    \Psi_{\text{inf}}(r) = \mathrm{e}^{\mathrm{i}\omega\, r_{\star}(r)}\sum^{\mathcal{I}}_{k=0}\mathfrak{p}_{k}\,r^{-k}.
\end{equation}
Here, $\{\mathfrak{h}_j\}$ and $\{\mathfrak{p}_k\}$ are the expansion coefficients at the horizon and infinity, respectively, which are typically functions of the black hole parameters and the frequency $\omega$. $\kappa=f'(r_{h})/2$ is the surface gravity at the event horizon,  $\mathcal{H}$ and $\mathcal{I}$ are the truncation orders.

To determine these coefficients $\{\mathfrak{h}_j\}$ and $\{\mathfrak{p}_k\}$, we substitute the series solutions, $\Psi_{\text{h}}(r)$ and $\Psi_{\text{inf}}(r)$ into Eq. (\ref{master_equation_frequency_domian}), i.e., $\mathcal{M}[\Psi_{\text{h}}(r)]$, $\mathcal{M}[\Psi_{\text{inf}}(r)]$. We then expand the resulting expressions around $r_h$ and $\infty$. By collecting terms of the same order, we transform the differential equation into two algebraic power series equations:
\begin{align}
    \sum^{\mathcal{H}}_{j=0}\mathcal{C}_{j}(\mathfrak{h}_0,\mathfrak{h}_1,\cdots,\mathfrak{h}_j;\omega)(r-r_h)^{j} = 0\, ,~~~~~
    \sum^{\mathcal{I}}_{k=0}\mathcal{B}_{k}(\mathfrak{p}_0,\mathfrak{p}_1,\cdots,\mathfrak{p}_k;\omega)\,r^{-k} = 0\, .
\end{align}

For these series equations to be valid for all values of $r$ in their respective regions of convergence, the coefficient of each power term must vanish independently. These conditions, i.e., $\mathcal{C}_{j}=0$ and $\mathcal{B}_{k}=0$ for all $j$ and $k$, allow us to derive the following recurrence relations for the expansion coefficients:
\begin{align}
    \mathfrak{h}_{j} &= \sum^{j-1}_{i=0} \Lambda_{ji}\,\mathfrak{h}_{i} \quad \text{for} \quad j=1,2,\cdots,\mathcal{H}\, , \label{eq:recurrence h} \\
    \mathfrak{p}_{k} &= \sum^{k-1}_{i=0} \Upsilon_{ki}\,\mathfrak{p}_{i} \quad \text{for} \quad k=1,2,\cdots,\mathcal{I}\, . \label{eq:recurrence inf}
\end{align}
The coefficients $\Lambda_{ji}$ and $\Upsilon_{ki}$ depend on the black hole parameters, namely $\Lambda_{ji}=\Lambda_{ji}[f^{(n)}(r_h),r_{h},\ell,s,\omega,\kappa]$ and $\Upsilon_{ki}=\Upsilon_{ki}[f^{(n)}(\infty),\ell,s,\omega]$. These relations show that each coefficient can be determined from the preceding ones. These two systems of linear equations can be intuitively expressed in the matrix form, where the coefficients form a lower triangular matrix:
\begin{align}
    \begin{pmatrix}
        \mathfrak{h}_1 \\
        \mathfrak{h}_2 \\
        \mathfrak{h}_3 \\
        \vdots \\
        \mathfrak{h}_{\mathcal{H}}
    \end{pmatrix}
    &=
    \begin{pmatrix}
        \Lambda_{10} & 0 & 0 & 0 & \cdots & 0 \\
        \Lambda_{20} & \Lambda_{21} & 0 & 0 & \cdots & 0 \\
        \Lambda_{30} & \Lambda_{31} & \Lambda_{32} & 0 & \cdots & 0 \\
        \vdots & \vdots  & \vdots  & \vdots & \ddots & \vdots \\
        \Lambda_{\mathcal{H}0} & \Lambda_{\mathcal{H}1} & \Lambda_{\mathcal{H}2} & \Lambda_{\mathcal{H}3} & \cdots & \Lambda_{\mathcal{H}\,\mathcal{H}-1}
    \end{pmatrix}
    \begin{pmatrix}
        \mathfrak{h}_0 \\
        \mathfrak{h}_1 \\
        \mathfrak{h}_2 \\
        \vdots \\
        \mathfrak{h}_{\mathcal{H}-1}
    \end{pmatrix}\, ,\label{eq:recurrence h2} \\
    \begin{pmatrix}
        \mathfrak{p}_1 \\
        \mathfrak{p}_2 \\
        \mathfrak{p}_3 \\
        \vdots \\
        \mathfrak{p}_{\mathcal{I}}
    \end{pmatrix}
    &=
    \begin{pmatrix}
        \Upsilon_{10} & 0 & 0 & 0 & \cdots & 0 \\
        \Upsilon_{20} & \Upsilon_{21} & 0 & 0 & \cdots & 0 \\
        \Upsilon_{30} & \Upsilon_{31} & \Upsilon_{32} & 0 & \cdots & 0 \\
        \vdots & \vdots  & \vdots  & \vdots & \ddots & \vdots \\
        \Upsilon_{\mathcal{I}0} & \Upsilon_{\mathcal{I}1} & \Upsilon_{\mathcal{I}2} & \Upsilon_{\mathcal{I}3} & \cdots & \Upsilon_{\mathcal{I}\,\mathcal{I}-1}
    \end{pmatrix}
    \begin{pmatrix}
        \mathfrak{p}_0 \\
        \mathfrak{p}_1 \\
        \mathfrak{p}_2 \\
        \vdots \\
        \mathfrak{p}_{\mathcal{I}-1}
    \end{pmatrix}\, . \label{eq:recurrence inf2}
\end{align}
Since the expressions for the higher-order coefficients $\Lambda_{ji}$ are very lengthy, we only show the first few $\Lambda_{ji}$ and $\Upsilon_{ki}$ here:
\begin{equation}\label{eq:recurrence coe}
    \begin{split}
        \Lambda_{10} &= \frac{1}{2 r_h^2 \left(\kappa ^2 \omega ^2 f'_{r_{h}}+\left(f'_{r_{h}}\right)^3 (\kappa -i \omega )^2\right)} \Big[ r_h \left(\left(f'_{r_{h}}\right)^3 \left(6 \omega ^2-2 \kappa ^2 (s-1)\right)-6 \kappa ^2 \omega ^2 f'_{r_{h}}\right) \\
        & \quad + 2 \kappa ^2 \ell (\ell +1) \left(f'_{r_{h}}\right)^2+f''_{r_{h}} r_h^2 \left(\kappa ^2 \omega ^2+\left(f'_{r_{h}}\right)^2 \left(2 \kappa ^2 \delta _{s-2}+\omega (\omega +i \kappa )\right)\right) \Big]\, , \\
        \\
        \Lambda_{20} &= \frac{1}{12 \left(f'_{r_{h}}\right)^2 r_h^3 \left(\kappa ^2 \omega ^2+\left(f'_{r_{h}}\right)^2 (2 \kappa -i \omega )^2\right)} \Big[ -3 \kappa ^2 \omega ^2 \left(f''_{r_{h}}\right)^2 r_h^3 - 36 \kappa ^2 \omega ^2 \left(f'_{r_{h}}\right)^2 r_h \\
        & \quad + 12 \left(f'_{r_{h}}\right)^4 r_h \left(3 \omega ^2-2 \kappa ^2 (s-1)\right) + 2 \kappa ^2 \omega ^2 f'_{r_{h}} r_h^2 \left(f^{(3)}_{r_{h}} r_h+9 f''_{r_{h}}\right) +6 \kappa ^2 \ell (\ell +1) \\
        & \quad + 2 \left(f'_{r_{h}}\right)^3 \left(f^{(3)}_{r_{h}} r_h^3 \left(6 \kappa ^2 \delta _{s-2}+\omega (\omega +2 i \kappa )\right)+3 f''_{r_{h}} r_h^2 \left(3 i \kappa \omega +\kappa ^2 \left(6 \delta _{s-2}-2 s+2\right)+3 \omega ^2\right)\right) \Big]\, , \\
        \\
        \Lambda_{21} &= \frac{1}{2 r_h^2 \left(\kappa ^2 \omega ^2 f'_{r_{h}}+\left(f'_{r_{h}}\right)^3 (2 \kappa -i \omega )^2\right)} \Big[ 2 f'_{r_{h}} \left(\kappa ^2 \ell (\ell +1) f'_{r_{h}} \right. \\
        & \quad \left. + r_h \left(-3 \kappa ^2 \omega ^2+\left(f'_{r_{h}}\right)^2 \left(6 i \kappa \omega -\left(\kappa ^2 (s+2)\right)+3 \omega ^2\right)\right)\right) \\
        & \quad + f''_{r_{h}} r_h^2 \left(\kappa ^2 \omega ^2+\left(f'_{r_{h}}\right)^2 \left(-2 \kappa ^2+3 i \kappa \omega +2 \kappa ^2 \delta _{s-2}+\omega ^2\right)\right) \Big]\, , \\
        \\
        \Upsilon_{10} &=\frac{i\,\ell  (\ell +1)}{2 \omega },\Upsilon_{20} =\frac{i f_{\infty }' \left(f_{\infty } \left(2 \delta _{s-2}+s-1\right)+\ell  (\ell +1)\right)}{4 \omega  f_{\infty }}\, ,~~
        \Upsilon_{21} =-\frac{f_{\infty }'}{2 f_{\infty }}+\frac{i \left(-2 f_{\infty }+\ell ^2+\ell \right)}{4 \omega }\, ,
    \end{split}
\end{equation}
in which we denote that $f_{r_h}^{(n)}=\partial^{(n)}_rf|_{r\rightarrow r_h}$ and $f_{\infty}^{(n)}=\partial^{(n)}_rf|_{r\rightarrow \infty}$. By solving the recurrence relations \eqref{eq:recurrence h} and \eqref{eq:recurrence inf} iteratively, we can express all higher-order expansion coefficients in terms of the leading-order coefficients, $\mathfrak{h}_0$ and $\mathfrak{p}_0$:
\begin{align}
    \mathfrak{h}_{j} &= \mathfrak{h}_{0}\, \mathsf{G}_{j}(\Lambda_{10},\cdots,\Lambda_{j\,j-1}) \quad \text{for} \quad j=1,2,\cdots,\mathcal{H}\, ,\\
    \mathfrak{p}_{k} &= \mathfrak{p}_{0}\, \mathsf{G}_{k}(\Upsilon_{10},\cdots,\Upsilon_{k\,k-1}) \quad \text{for} \quad k=1,2,\cdots,\mathcal{I}.
\end{align}
Due to the structural similarity of the linear systems \eqref{eq:recurrence h2} and \eqref{eq:recurrence inf2}, the general solutions for the coefficients at the horizon and at infinity share the same functional form, denoted by $\{\mathsf{G}_i\}$. However, the truncation orders, $\mathcal{H}$ and $\mathcal{I}$, can be differ. For clarity, we list the explicit forms of the first few functions $\{\mathsf{G}_i\}$:
\begin{eqnarray}\label{eq:h&p sol2}
    \begin{split}
        \mathsf{G}_{1} &= \mathcal{A}_{10}\, ,\mathsf{G}_{2} = \mathcal{A}_{20}+\mathcal{A}_{10} \mathcal{A}_{21}\, ,\mathsf{G}_{3} =\mathcal{A}_{30}+\mathcal{A}_{20} \mathcal{A}_{32}+\mathcal{A}_{10} \left(\mathcal{A}_{31}+\mathcal{A}_{21} \mathcal{A}_{32}\right)\, ,
        \\
        \mathsf{G}_{4} &=\mathcal{A}_{40}+\mathcal{A}_{20} \mathcal{A}_{42}+\left(\mathcal{A}_{30}+\mathcal{A}_{20} \mathcal{A}_{32}\right) \mathcal{A}_{43}+\mathcal{A}_{10} \left(\mathcal{A}_{41}+\mathcal{A}_{21} \mathcal{A}_{42}+\left(\mathcal{A}_{31}+\mathcal{A}_{21} \mathcal{A}_{32}\right) \mathcal{A}_{43}\right)\, ,
        \\
        \mathsf{G}_{5} &= \mathcal{A}_{50}+\mathcal{A}_{10} \mathcal{A}_{51}+\left(\mathcal{A}_{20}+\mathcal{A}_{10} \mathcal{A}_{21}\right) \mathcal{A}_{52}+\left(\mathcal{A}_{30}+\mathcal{A}_{20} \mathcal{A}_{32}+\mathcal{A}_{10} \left(\mathcal{A}_{31}+\mathcal{A}_{21} \mathcal{A}_{32}\right)\right) \mathcal{A}_{53}
        \\
        &\quad +\left(\mathcal{A}_{40}+\mathcal{A}_{20} \mathcal{A}_{42}+\left(\mathcal{A}_{30}+\mathcal{A}_{20} \mathcal{A}_{32}\right) \mathcal{A}_{43}+\mathcal{A}_{10} \left(\mathcal{A}_{41}+\mathcal{A}_{21} \mathcal{A}_{42}+\left(\mathcal{A}_{31}+\mathcal{A}_{21} \mathcal{A}_{32}\right) \mathcal{A}_{43}\right)\right) \mathcal{A}_{54}\, ,
    \end{split}
\end{eqnarray}
where $\mathcal{A}_{ij}$ represents either $\Lambda_{ij}$ (for the horizon expansion) or $\Upsilon_{ij}$ (for the infinity expansion).

Substituting these solutions back into the asymptotic expansions and setting $\mathfrak{h}_{0}=\mathfrak{p}_{0}=1$ without loss of generality, we obtain well-defined asymptotic forms of the radial function. Given $\omega\in\mathbb{C}$, the series $\Psi_{\text{h}}(\tilde{r}_h, \omega)$ and $\Psi_{\text{inf}}(\tilde{r}_{\infty}, \omega)$, provide the initial conditions for solving Eq. (\ref{master_equation_frequency_domian}). Here, $\tilde{r}_h=(1+\epsilon)r_{h}$ is a point slightly away from the horizon, with $\epsilon$ being a small parameter to avoid the singularity, while $\tilde{r}_{\infty}\gg r_{h}$ is a sufficiently large, which numerically represents infinity. In practice, for a given initial guess for frequency $\omega_g$, Eq. (\ref{master_equation_frequency_domian}) can be solved numerically as an initial value problem, starting from both boundaries:
\begin{align}
    \mathcal{M}[\Psi_{1}]&=0\, , \quad \text{with initial conditions at } \tilde{r}_h: \quad \Psi_{1}=\Psi_{\text{h}}(\tilde{r}_h,\omega_g)\, , \quad \Psi'_{1}=\Psi'_{\text{h}}(\tilde{r}_h,\omega_g)\, ,\\
    \mathcal{M}[\Psi_{2}]&=0\, , \quad \text{with initial conditions at } \tilde{r}_{\infty}: \quad \Psi_{2}=\Psi_{\text{inf}}(\tilde{r}_{\infty},\omega_g)\, , \quad \Psi'_{2}=\Psi'_{\text{inf}}(\tilde{r}_{\infty},\omega_g)\, .
\end{align}

This procedure yields two numerical solutions, $\Psi_{1}(r,\omega_{g})$ and $\Psi_{2}(r,\omega_{g})$, which are obtained by integrating outwards from the horizon and inwards from infinity, respectively. A true QNM spectum of Eq. (\ref{master_equation_frequency_domian}) is found only when these two solutions match smoothly. This matching condition is enforced by requiring their Wronskian to vanish at an arbitrary matching point $r_m$ (where $\tilde{r}_{h}<r_{m}<\tilde{r}_{\infty}$):
\begin{eqnarray}
    \mathcal{W}[\Psi_{1},\Psi_{2}] = \Psi_{1}(r_{m},\omega_{g})\Psi'_{2}(r_{m},\omega_{g})-\Psi'_{1}(r_{m},\omega_{g})\Psi_{2}(r_{m},\omega_{g})=0\, .
\end{eqnarray}
Therefore, we find the QNM spectra by treating the Wronskian as a function of the complex frequency $\omega_g$ and searching for its roots. We employ a numerical root-finding algorithm, such as Newton's method, to iteratively refine the guess $\omega_g$ until the Wronskian is zero to a desired precision. Once an eigenvalue $\omega_{\text{QNM}}$ is found, the corresponding global eigenfunction can be constructed by joining the two numerical solutions at the matching point, i.e.,
\begin{equation}\label{eigenfunction}
    \Psi(r,\omega_{\text{QNM}})=
    \begin{cases}
        \mathcal{N}_1 \Psi_{1}(r,\omega_{\text{QNM}}), & r \leq r_{m}, \\
        \mathcal{N}_2 \Psi_{2}(r,\omega_{\text{QNM}}), & r > r_{m},
    \end{cases}
\end{equation}
where $\mathcal{N}_1$ and $\mathcal{N}_2$ are normalization constants chosen to ensure continuity of the function at $r_m$. Note that in order to mitigate the influence of finite boundary effects on numerical results, we thus adopt higher truncation orders. In our numerical calculations for the GD hairy black hole, we use the truncation orders $\mathcal{H}=13$ and $\mathcal{I}=14$, with the integration boundaries set at $\epsilon=10^{-4}$ and $\tilde{r}_{\infty}=90M$.

\section{Methodology for ringdown analysis}
\label{sec:appendix_methodology}
In this appendix, we detail the numerical methodology used to generate the time-domain waveform and subsequently extract the QNM spectra and their corresponding excitation properties. Our approach is a multi-step process designed to first produce a reliable time-series signal and then robustly identify the mode frequencies and their physical contributions.

\subsection{Time-domain evolution: finite difference method}
\label{sec:appendix_finite difference}

To obtain the time-domain profile of the master scalar, we solve the master equation \eqref{eq: master eq} using the finite difference method in time profile. For more details on this standard approach, readers may refer to \cite{Zhu:2014sya, Yang:2023lcm, Yang:2024cjf}.

We begin by discretizing  $r_{\star}=j\Delta r_{\star}$ and time coordinate $t=i \,\Delta t$. The master scalar and the potential are evaluated at discrete grid points $\Psi(t,r_{\star})=\Psi(i\,\Delta t,j\,\Delta r_{\star})=\Psi_{i,j}$ and $V(r)=V(r(r_{\star}))=V(r(j\Delta r_{\star}))=V(r_j)=V_j$. The second-order time and space derivatives in the master equation are then approximated using central difference stencils. This transforms the partial differential equation into an iterative scheme that evolves the waveform in time:
\begin{eqnarray}\label{difference_stencils}
    \Psi _{i+1,j}=\frac{\Delta t^2 (\Psi _{i,j-1}-2 \Psi _{i,j}+\Psi _{i,j+1})}{ r_{\star}^2}-\Delta t^2 V_j \Psi _{i,j}+2 \Psi _{i,j}-\Psi _{i-1,j}\, .
\end{eqnarray}
To ensure the stability of this numerical scheme, the Courant-Friedrichs-Lewy (CFL) condition must be satisfied. For this wave-like equation, the condition requires $\Delta t / \Delta r_{\star} \leq 1$ \cite{Smith1985}. We choose our grid spacings to be $\Delta r_{\star} = 0.004$ and $\Delta t = 0.002$ to safely meet this criterion.

The evolution is initiated with a Gaussian wave packet of the form $\Psi(t=0, r_\star) = c \cdot \exp[-(r_{\star}- a)^2 / (2b^2)]$, with the field being zero for $t<0$. The resulting waveform typically exhibits three distinct phases: an initial outburst from the source, the characteristic damped oscillations of the ringdown phase, and a late-time power-law tail. It is from the ringdown phase that we extract the QNMs and their complex amplitudes using methods described below.

\subsection{QNM spectra extraction: matrix pencil method}
\label{sec:appendix_Matrix pencil}
To determine the signal poles $\{z_k\}$, we employ the Matrix Pencil (MP) method, a robust technique known for its high accuracy and stability in the presence of noise~\cite{Berti:2007dg,56027, 370583}. This method circumvents the potential instabilities of traditional Prony methods by recasting the problem in terms of a generalized eigenvalue problem.

The procedure begins with constructing two $(N-L) \times L$ Hankel matrices, $Y_0$ and $Y_1$, from the $N$ data points of the time series $x[n]$. $L$ is the pencil parameter, chosen to be in the range $p \le L \le N-p$. The matrices are defined as:
\begin{equation}
    Y_0 =
    \begin{pmatrix}
        x[0] & x[1] & \cdots & x[L-1] \\
        x[1] & x[2] & \cdots & x[L] \\
        \vdots & \vdots & \ddots & \vdots \\
        x[N-L-1] & x[N-L] & \cdots & x[N-2]
    \end{pmatrix},
\end{equation}
and
\begin{equation}
    Y_1 =
    \begin{pmatrix}
        x[1] & x[2] & \cdots & x[L] \\
        x[2] & x[3] & \cdots & x[L+1] \\
        \vdots & \vdots & \ddots & \vdots \\
        x[N-L] & x[N-L+1] & \cdots & x[N-1]
    \end{pmatrix}.
\end{equation}
These matrices are then used to form a matrix pencil, $Y_1 - zY_0$. The signal poles $\{z_k\}$ are the generalized eigenvalues of the matrix pair $(Y_1, Y_0)$ that reduce the rank of the pencil. Numerically, this is achieved by first performing a Singular Value Decomposition (SVD) via $Y_0=USV^{T}$ and then estimating the largest singular values of $S$ as our mode order $p$ . The poles $\{z_k\}$ are then found as the standard eigenvalues of a reduced $p \times p$ matrix $Z_p=S^{-1}_{p}U^{T}_{p}Y_{0}V_{p}$ constructed from the SVD components of $Y_0$ and the Hankel matrix $Y_1$. Our QNM data is extracted by setting pencil parameter $L=N/2$.

\subsection{Complex amplitude determination}
\label{sec:appendix_least-squares}
With the set of poles $\{z_k\}$ determined, we find the complex amplitudes $\{h_k\}$ by solving the linear system defined by Eq. \eqref{eq:signal_model} using all $N$ available data points. This forms an overdetermined system of linear equations:
\begin{equation}
    \begin{pmatrix}
        z_1^0 & z_2^0 & \cdots & z_p^0 \\
        z_1^1 & z_2^1 & \cdots & z_p^1 \\
        \vdots & \vdots & \ddots & \vdots \\
        z_1^{N-1} & z_2^{N-1} & \cdots & z_p^{N-1}
    \end{pmatrix}
    \begin{pmatrix}
        h_1 \\
        h_2 \\
        \vdots \\
        h_p
    \end{pmatrix}
    =
    \begin{pmatrix}
        x[0] \\
        x[1] \\
        \vdots \\
        x[N-1]
    \end{pmatrix}.
    \label{eq:vandermonde}
\end{equation}
This system, in the form $\mathbf{V}\mathbf{h} = \mathbf{x}$, where $\mathbf{V}$ is a Vandermonde matrix, is solved for $\mathbf{h}$ using a linear least-squares algorithm. This ensures that the solution for $\{h_k\}$ is optimal for the entire signal segment under consideration.

\bibliography{reference}
\bibliographystyle{utphys}

\end{document}